\documentclass[a4paper,11pt]{article}
\usepackage{cite}
\usepackage{draft}
\usepackage{mathtools}
\usepackage{bm}
\usepackage{simplewick}
\usepackage{booktabs}
\usepackage{array}
\usepackage{microtype}
\usepackage{orcidlink}
\usepackage{enumitem}
\usepackage{float}
\usepackage[nameinlink,capitalise,noabbrev]{cleveref}
\usepackage{tikz}
\usetikzlibrary{arrows.meta,positioning,calc}
\usepackage[utf8]{inputenc}
\usepackage{textalpha}
\crefformat{section}{\S#2#1#3} 
\crefformat{subsection}{\S#2#1#3}
\crefformat{subsubsection}{\S#2#1#3}
\definecolor{arrowgreen}{RGB}{74,145,105}
\definecolor{lightgreenbox}{RGB}{232,246,237}
\definecolor{lightbluebox}{RGB}{232,241,250}
\definecolor{lightgoldbox}{RGB}{252,245,224}
\definecolor{lightredbox}{RGB}{250,232,235}
\definecolor{boxborder}{RGB}{75,95,105}
\usepackage{verbatim}

\definecolor{linkblue}{RGB}{25, 70, 120}
\definecolor{citegreen}{RGB}{35, 105, 80}
\definecolor{urlteal}{RGB}{20, 105, 115}
\definecolor{fileburgundy}{RGB}{125, 45, 65}

\hypersetup{
    breaklinks=true,
    colorlinks=true,
    linkcolor=linkblue,
    citecolor=citegreen,
    urlcolor=urlteal,
    filecolor=fileburgundy
}

\hypersetup{
  pdftitle={BRST quantization of Carroll-Weyl gauged null strings},
  pdfauthor={Sarthak Duary and Sourav Maji},
  pdfsubject={BRST quantization of the Carroll-Weyl bosonic null string},
  pdfkeywords={BRST quantization, null strings, Carroll-Weyl symmetry, flipped vacuum, BMS symmetry, operator product expansions, critical dimension}
}

\allowdisplaybreaks[3]
\usepackage{array,tabularx}
\usepackage{float}
\newcommand{\dd}{\mathrm{d}}
\newcommand{\ii}{\mathrm{i}}
\newcommand{\e}{\mathrm{e}}
\newcommand{\cL}{c_{LL}}
\newcommand{\cS}{c_{LS}}
\newcommand{\cSS}{c_{SS}}

\newcommand{\Res}{\mathop{\mathrm{Res}}}

\newcommand{\cH}{\mathcal{H}}

\newcommand{\gh}{\mathrm{gh}}
\newcommand{\mat}{\mathrm{mat}}
\newcommand{\tot}{\mathrm{tot}}
\newcommand{\CW}{\mathrm{CW}}
\newcommand{\BMS}{\mathrm{BMS}}

\newcommand{\half}{\tfrac{1}{2}}
\newcommand{\normal}[1]{:\!#1\!:}

\title{BRST quantization of Carroll-Weyl gauged null strings}

\author[a]{Sarthak Duary\,\orcidlink{0000-0002-4535-3198}}
\author[b]{and Sourav Maji\,\orcidlink{0009-0008-2963-2497}}

\affiliation[a]{Yau Mathematical Sciences Center (YMSC),
  Tsinghua University, Beijing 100084, China}

\affiliation[b]{Harish-Chandra Research Institute,
  A CI of Homi Bhabha National Institute,
  Chhatnag Road, Jhunsi, Prayagraj (Allahabad),
  Uttar Pradesh 211019, India}

\emailAdd{sarthakduary@tsinghua.edu.cn}
\emailAdd{souravmaji@hri.res.in}

\abstract{We study the BRST quantization of the null string after completing its local gauge symmetry by Carroll-Weyl transformations. The resulting worldsheet theory possesses three first-class constraints, $C_1 = P^2$, $C_2 = P \cdot X'$, and $C_3 = P \cdot X$, whose modes realize a Weyl-BMS algebra. The additional Carroll-Weyl constraint qualitatively changes the quantum gauge complex: its scalar $s$-ghost is intrinsically coupled to the BMS $bc$-ghost sector, and the anomaly analysis involves three independent cocycles rather than a single Virasoro-type central charge. Starting from the gauge-fixed action, we derive the complete Faddeev-Popov complex, construct the matter and ghost currents and the BRST charge, and evaluate their equal-time operator products in the flipped, equivalently highest-weight, representation. The matter and ghost anomaly coefficients are $(c_{LL},c_{LS},c_{SS})_{\mathrm{matter}}=(2D,-D,-D)$ and $(c_{LL},c_{LS},c_{SS})_{\mathrm{ghost}}=(-54,6,4)$. Because the corresponding central terms multiply linearly independent ghost bilinears in $Q_B^2$, BRST nilpotency requires the three conditions $D=27$, $D=6$, and $D=4$, respectively. These conditions are mutually incompatible. Consequently, there is no target-space dimension in which the minimal flat Carroll-Weyl matter-plus-ghost complex is anomaly-free in the highest-weight representation. The familiar $D=26$ condition of the ILST null string is recovered only after truncation to the two-constraint BMS subsector, which defines a different quantum gauge complex.}

\begin{document}
\maketitle

\section{Introduction}
\label{sec:intro}

BRST quantization provides the natural framework for determining
whether a gauge theory admits a consistent quantum description. Its
construction requires the complete set of local gauge symmetries, the
introduction of a corresponding ghost system for each independent
generator, and the formation of the total quantum constraint algebra.
For the tensile bosonic string, conformal gauge leads to the
fermionic \((b,c)\) ghosts, and the cancellation between the matter
and ghost central charges ensures the nilpotency of the BRST
operator, yielding \(D = 26\) as a consequence
\cite{Kato:1983,Polchinski1}. Thus, the critical dimension is not the
primary input of the quantization procedure, nor is it a property of
the matter kinetic term alone; rather, it emerges from the
requirement that the full BRST complex be anomaly free. A BRST
construction based only on a proper subalgebra of the gauge
symmetries may define a mathematically consistent cohomological
problem, but it does not necessarily constitute a quantization of the
original gauge theory.

Null strings are among the earliest and most direct realizations of Carroll symmetry, the ultra-relativistic contraction associated with a vanishing speed of light \cite{Schild1977,Isberg1994}. Their worldsheet metric is
degenerate, and after the conventional temporal gauge fixing the
residual symmetry is the three-dimensional Bondi-Metzner-Sachs algebra \(\BMS_3\) \cite{Bondi:1962px,Sachs1962AsymptoticSI,Barnich:2006av}, equivalently the
two-dimensional Galilean conformal algebra
\cite{Bagchi2013,BagchiChakraborttyParekh2016}.  The same algebraic
structure also lies behind the null origin of ambitwistor strings and
their connection with scattering equations
\cite{MasonSkinner2014,CasaliTourkine2016,CasaliHerfrayTourkine2017}.
These links make the quantum null string relevant well beyond the
formal zero-tension limit.  They also sharpen a long-standing warning:
different normal orderings and inequivalent vacua can lead to different
critical-dimension statements
\cite{Lizzi1986,Bozhilov1997,BagchiVacua2021}.  A quantum result is
therefore meaningful only after both the gauge complex and its
representation have been specified.

A complementary perspective is provided by the recently quantized
Carrollian bosonic string of \cite{FigueroaHaveObers2025}.  In that theory both the worldsheet and
the target spacetime are Carrollian.  Its residual \(\BMS_3\) symmetry
admits a BRST treatment in terms of a formal Laurent variable, and its
cohomology organizes a finite spectrum into representations of the
Carroll group.  The model studied here is different: our target
embedding is flat, while the worldsheet is null and the local gauge
orbit is completed by a Carroll-Weyl symmetry \cite{SheikhJabbari2026a,SheikhJabbari2026b}.  Nevertheless, the two
problems share an important lesson.  BMS-type mode algebras can be
treated cohomologically, but only after the currents, ghosts, and
meaning of the Laurent expansion have been derived from the underlying
worldsheet theory.  Recent work on chiral BMS-like algebras supplies
the corresponding semi-infinite and BRST machinery
\cite{FigueroaVishwa2025,Batlle2024}.

The conventional ILST null string retains two first-class constraints:
a spatial diffeomorphism and a null supertranslation.  Its
Faddeev-Popov determinant therefore contains two ghost pairs.  In the
flipped, equivalently highest-weight, representation, this
two-constraint system reproduces a \(D=26\) condition
\cite{Chen2023,BagchiVacua2021}.  This is a consistent statement about
that reduced gauge complex.  The central question of the present paper
is whether it remains true once the full local symmetry of the
degenerate worldsheet is included.

The answer requires a third constraint.  A volume-preserving
Carroll-Weyl transformation rescales the embedding and the Carroll
vector while shifting a Weyl connection \(W_a\) \cite{SheikhJabbari2026b}.  Its Hamiltonian
moment map completes the constraint set to
\begin{equation}
 C_1=P^2,\qquad
 C_2=P\cdot X',\qquad
 C_3=P\cdot X .
\label{eq:intro-third-constraint}
\end{equation}
These are the conventions of our preceding path-integral analysis
\cite{Duary:2026bcs}.  For the operator-product calculation it is
convenient to pass to
\begin{equation}
 M=\frac12C_1,\qquad L=-C_2,\qquad S=C_3 .
\end{equation}
The modes form a Weyl-BMS semidirect product, with the characteristic
bracket
\begin{equation}
 [S_m,M_n]=2M_{m+n}.
\end{equation}
The underlying symmetry and its classical consequences were identified
in \cite{SheikhJabbari2026a,SheikhJabbari2026b,SheikhJabbari2026c};
its canonical Weyl-BMS realization is closely related to
\cite{Batlle2020}. Related diffeomorphism-Weyl structures also arise in gravitational null-boundary charge algebras \cite{Adami:2020ugu,Adami:2021nnf}. The third first-class constraint changes the local
configuration-space count from \(D-2\) to \(D-3\).  At the quantum
level it does more: it introduces a scalar ghost pair, opens mixed
\(LS\) and affine \(SS\) anomaly channels, and couples the
supertranslation and Carroll-Weyl sectors.  Imposing \(C_3\) only after
quantizing the two-constraint system cannot recover these effects.

The three-row Faddeev-Popov operator derived in
\cite{Duary:2026bcs} makes this coupling concrete.  Its determinant
produces two weight-\((2,-1)\) ghost pairs and a weight-\((1,0)\)
Carroll-Weyl pair.  The gauge-fixed cylinder action also contains the
algebraic term \(-2sb_0\).  This term is not an incidental interaction:
it is the Lagrangian image of \([S,M]=2M\), and it forces the scalar
ghost to participate in a genuinely semidirect \(bcs\) complex.  A
calculation that simply adds the central charge of an independent
weight-\((1,0)\) system therefore misses the mixed ghost currents and
cannot decide BRST nilpotency.

Our purpose is to construct this gauge-complete quantum complex from
first principles and to test it without assuming the answer.  The
calculation proceeds on a fixed Carrollian-time slice.  Following the
formal-variable logic used for BMS-type theories
\cite{FigueroaHaveObers2025,FigueroaVishwa2025}, the coordinate
\(z=\e^{\ii\sigma}\) packages Fourier modes on the spatial circle into
Laurent fields.  It is not an additional worldsheet coordinate and
does not imply a relativistic holomorphic factorization.  The
flipped-vacuum (highest-weight) splitting specifies the contractions,
while the coupled equations of motion reconstruct the unequal-time
fields.  The full two-coordinate Carrollian lift contains descendants
of the same equal-time algebra and introduces no additional
independent central data \cite{Hao2022}.

\paragraph{Strategy.}
The logical structure of the computation is
\begin{equation}
\begin{gathered}
 S_{\rm tensile}
 \xrightarrow{\ \mathcal T\to0\ }
 S_{\rm ILST}
 \xrightarrow{\ \text{Carroll-Weyl completion}\ }
 S_{\CW}[X,V,W]
 \xrightarrow{\ \text{gauge fixing}\ }
 S_X+S_{bcs}
 \\[4pt]
 \Downarrow
 \\[-2pt]
 (C_2,C_1,C_3)
 \longrightarrow
 (T^{\tot},M^{\tot},S^{\tot})
 \xrightarrow{\ \text{equal-time OPEs}\ }
 (\cL,\cS,\cSS)_{\tot}
 \xrightarrow{\ \text{BRST}\ }
 Q_B^{\,2}.
\end{gathered}
\label{eq:paper-concept-map}
\end{equation}
Every arrow in eq.~\eqref{eq:paper-concept-map} is part of the
derivation.  In particular, the matter currents are obtained as moment
maps of the Carroll-Weyl gauged action, and the ghost stress tensors
and mixed currents follow from the Faddeev-Popov action and the
Weyl-BMS structure constants.  Only after these steps do we compute
the singular operator products.  This order separates three logically
distinct questions: completeness of the classical gauge system,
closure of its quantum current algebra, and nilpotency of the BRST
charge.

\paragraph{Main Results.}
The matter currents are
\begin{equation}
 \mathcal C_1^X=\normal{P^2},\qquad
 \mathcal C_2^X=\normal{P\cdot\partial X},\qquad
 \mathcal C_3^X=\normal{X\cdot P},
\end{equation}
or, in the OPE-adapted basis,
\begin{equation}
 M^X=\frac12\mathcal C_1^X,\qquad
 T^X=-\mathcal C_2^X,\qquad
 S^X=\mathcal C_3^X.
\end{equation}
Single contractions reproduce the centerless Weyl-BMS brackets.
Double contractions give the matter anomaly vector
\begin{equation}
 (\cL,\cS,\cSS)_{\mat}=(2D,-D,-D).
\end{equation}
The complete ghost currents yield
\begin{equation}
 (\cL,\cS,\cSS)_{\gh}=(-54,6,4),
\end{equation}
and hence
\begin{equation}
 \cL^{\tot}=2D-54,\qquad
 \cS^{\tot}=6-D,\qquad
 \cSS^{\tot}=4-D .
\label{eq:intro-result}
\end{equation}
These coefficients are not three ways of reporting one Virasoro
anomaly.  The \(S_0\) Jacobi identity removes an independent \(LM\)
cocycle \cite{Batlle2024}, while the surviving \(LL\), \(LS\), and \(SS\) cocycles
multiply linearly independent \(cc\), \(cs\), and \(ss\) ghost
bilinears in \(Q_B^2\).  BRST nilpotency therefore requires all three
entries of eq.~\eqref{eq:intro-result} to vanish separately.  They
would select \(D=27\), \(D=6\), and \(D=4\), respectively, and thus
have no common solution.  The isolated value \(D=27\) cancels only the
Virasoro-type cocycle; it is not a critical dimension of the complete
theory.

The conclusion is precise: in the flipped-vacuum representation, there is no target-space dimension for which the BRST charge of the minimal flat Carroll-Weyl matter-plus-ghost complex is nilpotent. This
does not rule out other quantum theories.  An induced representation
changes the contraction prescription, a reduced light-cone
quantization removes the unreduced BRST algebra, and additional matter
or non-minimal fields can alter the anomaly vector
\cite{Rasulian2026}.  Our result instead identifies exactly why the
single \(D=26\) test of the two-constraint BMS string does not survive
the Carroll-Weyl completion.

\paragraph{Organization of the paper.}
\Cref{sec:tensile-brst-review} reviews the tensile BRST construction
and fixes the benchmark for the nilpotency test.
\Cref{sec:from-tensile} takes the tensionless limit, constructs the
Carroll-Weyl completion, derives the three-row Faddeev-Popov complex,
and defines the equal-time representation.  \Cref{sec:ope} obtains the
matter and ghost currents from the action and evaluates their anomaly
coefficients.  \Cref{sec:brst} constructs the quantum BRST charge and
proves that the three central obstructions must vanish independently.
\Cref{sec:interpret} compares this result with the two-constraint,
induced-vacuum, and reduced quantizations, and
\cref{sec:conclusion} discusses its implications and possible
extensions. \Cref{app:cylinder} develops the cylinder mode and Laurent-field dictionary, \Cref{app:ghost} derives the ghost currents from the BFV construction, \Cref{app:modes} analyzes the mode algebra, central extensions, and BRST obstruction, \Cref{app:BRST-current-transformation} presents an equivalent local OPE test of BRST consistency by examining the tensorial transformation of the BRST current, and \Cref{app:two-coordinate} extends the equal-time algebra to the full two-coordinate Carrollian plane.

\medskip
\noindent \textit{Note added: While finalizing this work, we became aware of the independent work of Bin Chen and Zezhou Hu~\cite{Chen:2026cau}, which appeared simultaneously with ours. They study quantum anomalies in four tensionless bosonic-string models by an independent method, viz.\ within a unified Hamiltonian and mode-algebra framework. In particular, their flipped-vacuum analysis of the Carroll-Weyl gauged null string finds the same three incompatible conditions, \(D=27\), \(D=6\), and \(D=4\), giving a cross-check.}

\section{Preliminaries: BRST quantization of the tensile bosonic string}
\label{sec:tensile-brst-review}

We begin with the ordinary closed bosonic string because its BRST
quantization supplies both the conceptual model and the calculational
standard for the Carroll-Weyl problem.  The presentation follows the
logic of \cite{Polchinski1}. We work in one holomorphic sector. The antiholomorphic
sector is an identical commuting copy, and the closed-string charge is
their sum.

\subsection{Gauge-fixed action, elementary OPEs, and stress tensors}

In conformal gauge the Euclidean action separates into matter and
Faddeev-Popov parts,
\begin{equation}
 S_{\rm tens}=S_X+S_{bc},\qquad
 S_X=\frac{1}{2\pi\alpha'}\int\dd^2z\,
 \partial X^\mu\bar\partial X_\mu,\qquad
 S_{bc}=\frac{1}{2\pi}\int\dd^2z\,b\,\bar\partial c .
\label{eq:tensile-gf-action}
\end{equation}
The equations of motion make \(b,c,\partial X^\mu\) holomorphic.  Our
elementary contractions are
\begin{align}
 X^\mu(z,\bar z)X^\nu(w,\bar w)
 &\sim-\frac{\alpha'}{2}\eta^{\mu\nu}\ln|z-w|^2,
\label{eq:tensile-XX}\\
 b(z)c(w)&\sim\frac{1}{z-w},\qquad
 c(z)b(w)\sim\frac{1}{z-w}.
\label{eq:tensile-bc}
\end{align}
The matter and ghost stress tensors are
\begin{align}
 T^X&=-\frac1{\alpha'}\normal{\partial X^\mu\partial X_\mu},
\label{eq:tensile-TX}\\
 T^{bc}&=-2\normal{b\partial c}-\normal{(\partial b)c}.
\label{eq:tensile-Tbc}
\end{align}
For example, either \(\partial X\) in \(T^X\) may contract with
\(X^\mu\), giving
\begin{equation}
 T^X(z)X^\mu(w,\bar w)\sim\frac{\partial X^\mu(w)}{z-w}.
\label{eq:tensile-TX-X}
\end{equation}
The fermionic Wick theorem similarly gives
\begin{align}
 T^{bc}(z)b(w)&\sim\frac{2b(w)}{(z-w)^2}
 +\frac{\partial b(w)}{z-w},\nonumber\\
 T^{bc}(z)c(w)&\sim-\frac{c(w)}{(z-w)^2}
 +\frac{\partial c(w)}{z-w}.
\label{eq:tensile-Tbc-fields}
\end{align}
Thus \(b,c\) have weights \(2,-1\) respectively.  Double contractions yield
\begin{align}
 T^X(z)T^X(w)&\sim
 \frac{D/2}{(z-w)^4}
 +\frac{2T^X(w)}{(z-w)^2}
 +\frac{\partial T^X(w)}{z-w},\nonumber\\
 T^{bc}(z)T^{bc}(w)&\sim
 -\frac{13}{(z-w)^4}
 +\frac{2T^{bc}(w)}{(z-w)^2}
 +\frac{\partial T^{bc}(w)}{z-w}.
\label{eq:tensile-central-opes}
\end{align}
Consequently \(c_X=D\), \(c_{bc}=-26\), and
\begin{equation}
 c_{\rm tot}=D-26.
\label{eq:tensile-ctot}
\end{equation}
At this point \(D=26\) is suggested by conformal-anomaly cancellation,
but the BRST calculation explains why the same number is required for a
well-defined gauge cohomology.

\subsection*{Classical BRST transformations}

We remove the constant odd parameter and denote the odd differential by
\(d_B\). The transformations in one chiral sector are
\begin{equation}
 d_BX^\mu=c\partial X^\mu,\qquad
 d_Bc=c\partial c,\qquad
 d_Bb=T^X+T^{bc}\equiv T^{\rm tot}.
\label{eq:tensile-brst-transformations}
\end{equation}
They are the gauge-fixed form of a worldsheet diffeomorphism with its
parameter replaced by the ghost.  The first two nilpotency checks are
immediate.  With the graded Leibniz rule,
\begin{align}
 d_B^2c
 &=(c\partial c)\partial c-c\,\partial(c\partial c)=0,
\label{eq:tensile-d2c}\\
 d_B^2X^\mu
 &=(c\partial c)\partial X^\mu
 -c\,\partial(c\partial X^\mu)=0,
\label{eq:tensile-d2X}
\end{align}
because \(c^2=(\partial c)^2=0\).  For \(b\), the statement
\(d_B^2b=d_BT^{\rm tot}=0\) holds classically on the gauge-fixed
equations of motion.  Quantum mechanically it is precisely this last
identity that can fail through the Virasoro central term.

\subsection{Construction of the BRST current}

The holomorphic current of ghost number one and weight one is
\begin{equation}
 j_B
 =\normal{cT^X}+\normal{bc\partial c}
 +\frac32\,\partial^2c.
\label{eq:tensile-jB}
\end{equation}
The first term inserts the matter constraint.  The cubic term is the
BFV term dictated by the Virasoro structure constants.  The improvement
\(\frac32\partial^2c\) is a total derivative and therefore does not
change the integrated charge on a closed contour, but it fixes the local
OPEs and makes the critical current transform as a weight-one tensor.
Equivalently, up to a total derivative,
\begin{equation}
 j_B=\normal{c\left(T^X+\frac12T^{bc}\right)}
 +\frac32\,\partial^2c.
\label{eq:tensile-jB-symmetric}
\end{equation}
The charge is
\begin{equation}
 Q=\oint\frac{\dd z}{2\pi\ii}\,j_B(z),
\qquad
 Q_B=Q+\bar Q
\label{eq:tensile-Q}
\end{equation}
for the closed string.

\subsection*{OPEs with \texorpdfstring{\(X^\mu,c,b\)}{X, c, b} and the antighost identity}

Only \(T^X\) contracts with \(X^\mu\), and only \(b\) in the cubic
term contracts with \(c\).  Hence
\begin{equation}
 j_B(z)X^\mu(w)\sim\frac{c\partial X^\mu(w)}{z-w},
\qquad
 j_B(z)c(w)\sim\frac{c\partial c(w)}{z-w}.
\label{eq:tensile-jB-Xc}
\end{equation}
Their residues reproduce the first two transformations in
eq.~\eqref{eq:tensile-brst-transformations}.

The antighost OPE contains more structure.  The three terms of
eq.~\eqref{eq:tensile-jB} give, respectively,
\begin{align}
 \normal{cT^X}(z)b(w)&\sim\frac{T^X(w)}{z-w},
\nonumber\\
\normal{ \frac32\partial^2c}(z)b(w)&\sim\frac{3}{(z-w)^3},
\nonumber\\
 \normal{bc\partial c}(z)b(w)
 &\sim\frac{j_{\rm gh}(w)}{(z-w)^2}
 +\frac{T^{bc}(w)}{z-w},
\qquad j_{\rm gh}=-\normal{bc}.
\label{eq:tensile-jB-b-pieces}
\end{align}
In the last line the simple pole follows by Taylor expanding the
uncontracted fields
\[
 \partial j_{\rm gh}-\normal{b\partial c}
 =-\normal{(\partial b)c}-2\normal{b\partial c}=T^{bc}.
\]
Adding the pieces,
\begin{equation}
 j_B(z)b(w)\sim
 \frac{3}{(z-w)^3}
 +\frac{j_{\rm gh}(w)}{(z-w)^2}
 +\frac{T^{\rm tot}(w)}{z-w}.
\label{eq:tensile-jB-b}
\end{equation}
Only the simple pole contributes to the zero-mode contour, so
\begin{equation}
 \{Q,b(w)\}=T^{\rm tot}(w).
\label{eq:tensile-antighost-identity}
\end{equation}
This identity is the local prototype for the three Carroll-Weyl
relations \(\{Q_B,b_A\}=G_A^{\rm tot}\) derived below.

\subsection*{Oscillator charge and ordering constant}

We now use
\begin{equation}
 T^X(z)=\sum_nL_n^Xz^{-n-2},\quad
 b(z)=\sum_nb_nz^{-n-2},\quad
 c(z)=\sum_nc_nz^{-n+1},\quad
 \{b_m,c_n\}=\delta_{m+n,0}.
\label{eq:tensile-modes}
\end{equation}
The residue of \(\normal{cT^X}\) gives
\(\sum_nc_{-n}L_n^X\).  Expanding the cubic term and
antisymmetrizing its two ghost indices gives the Virasoro structure
constant
\begin{equation}
 Q=
 \sum_n c_{-n}L_n^X
 -\frac12\sum_{m,n}(m-n)
 \normal{c_{-m}c_{-n}b_{m+n}}
 -a\,c_0 .
\label{eq:tensile-Q-modes}
\end{equation}
The sign of \(a\) follows the displayed convention for \(L_0-a\).
The local improvement integrates to zero because the residue selects
the coefficient \(n(n-1)c_n\) at \(n=0\).  Normal ordering of the
infinite cubic sum nevertheless permits the separate zero-mode term
\(-ac_0\).  Direct anticommutation gives
\begin{equation}
 \{Q,b_n\}=L_n^{\rm tot}-a\,\delta_{n,0}.
\label{eq:tensile-Qbn}
\end{equation}
Thus the intercept affects the zero-mode constraint, but it cannot
change the coefficient of the \(m^3\) Virasoro cocycle.

\subsection*{The \texorpdfstring{\(j_Bj_B\)}{BRST-current} OPE, nilpotency, and
\texorpdfstring{\(D=26\)}{D=26}}

The quantum consistency test is the self-OPE of
eq.~\eqref{eq:tensile-jB}.  Writing \(\delta=z-w\), the complete singular
part obtained from the matter-matter, matter-ghost, pure-ghost, and
improvement contractions is
\begin{align}
 j_B(z)j_B(w)\sim{}&
 -\frac{D-18}{2\delta^3}\,c\partial c(w)
 -\frac{D-18}{4\delta^2}\,c\partial^2c(w)
\nonumber\\
 &-\frac{D-26}{12\delta}\,c\partial^3c(w).
\label{eq:tensile-jBjB}
\end{align}
This formula also shows why it is not enough to inspect a single pole
before performing the contour integral.  The third-order pole gives
\(\oint\partial^2(c\partial c)\), and the second-order pole gives
\(\oint\partial(c\partial^2c)\); both are total derivatives on a
closed contour.  The simple pole is different
\begin{equation}
 Q^2=-\frac{D-26}{24}
 \oint\frac{\dd w}{2\pi\ii}\,c\,\partial^3c(w).
\label{eq:tensile-Q2}
\end{equation}
The operator under the contour is not identically zero.  Therefore
\begin{equation}
 Q_B^2=0\quad\Longleftrightarrow\quad D=26
\label{eq:tensile-critical-D}
\end{equation}
after imposing the corresponding condition in both closed-string
sectors.  This is the cohomological meaning of criticality:
\(\operatorname{im}Q_B\subset\ker Q_B\) only at the anomaly-free value.

\subsection*{The BRST current as a tensor}

There is an equivalent local test.  With
\(T^{\rm tot}=T^X+T^{bc}\), direct Wick contraction gives
\begin{equation}
 T^{\rm tot}(z)j_B(w)\sim
 \frac{D-26}{2(z-w)^4}\,c(w)
 +\frac{j_B(w)}{(z-w)^2}
 +\frac{\partial j_B(w)}{z-w}.
\label{eq:tensile-TjB}
\end{equation}
The unwanted fourth-order pole vanishes exactly when \(D=26\).
Hence three statements are equivalent
\begin{equation}
 c_{\rm tot}=0
 \quad\Longleftrightarrow\quad
 Q_B^2=0
 \quad\Longleftrightarrow\quad
 j_B\ \hbox{is a weight-one tensor}.
\label{eq:tensile-three-tests}
\end{equation}


The tensile derivation has a definite order: identify all constraints;
derive one ghost pair for each; construct the total constraint currents;
obtain the BRST transformations and current; verify the antighost
identities; compute every allowed central OPE; and only then test the
self-OPE or square of the charge.  The Carroll-Weyl theory differs at
three places: it has three constraints rather than one constraint per
relativistic chirality, its algebra is a semidirect Weyl-BMS algebra
rather than Virasoro, and its anomaly is a vector of independent
cocycles rather than a single central charge.  

\section{Carroll-Weyl gauged null strings: action, constraints, and gauge fixing}
\label{sec:from-tensile}
We now turn to the gauge structure of the Carroll-Weyl null string and identify the precise system to be quantized. We begin with the phase-space tensionless limit, which yields the familiar two-constraint null-string theory, and then incorporate the additional local Carroll-Weyl symmetry, whose moment map supplies a third first-class constraint. With the complete constraint system in hand, we fix the gauge and derive the corresponding Faddeev-Popov complex. We finally formulate the equal-time Laurent-field and OPE prescription that will be used in the quantum current-algebra analysis of the following section.


\subsection{Phase-space tensionless limit}

The most transparent route to the null string begins before conformal
gauge is imposed.  A bosonic string of tension \(\mathcal T\) admits the
first-order action
\begin{equation}
 S_{\mathcal T}
 =\frac{1}{2\pi}\int\dd\tau\,\dd\sigma
 \left[
 P\cdot\dot X
 -\frac{e}{2}\left(P^2+\mathcal T^2X'^2\right)
 -u\,P\cdot X'
 \right].
\label{eq:tensile-phase-space}
\end{equation}
The multipliers \(e\) and \(u\) impose the Hamiltonian and momentum
constraints
\begin{equation}
 \mathcal H_\perp
 =\frac12\left(P^2+\mathcal T^2X'^2\right)\approx0,
 \qquad
 \mathcal H_\parallel=P\cdot X'\approx0.
\label{eq:tensile-constraints}
\end{equation}
At nonzero tension these are two presentations of the two components
of the worldsheet stress tensor.  Their Poisson brackets contain
\(\mathcal T^2\), and after the usual linear combinations they become
two commuting classical Virasoro algebras.

Taking \(\mathcal T\to0\) at fixed \(P,e,u\) gives
\begin{equation}
 S_{0}
 =\frac{1}{2\pi}\int\dd\tau\,\dd\sigma
 \left(P\cdot\dot X-\frac e2P^2-uP\cdot X'\right),
\label{eq:null-phase-space}
\end{equation}
with constraints
\begin{equation}
 C_1=P^2\approx0,\qquad C_2=P\cdot X'\approx0.
\label{eq:old-null-constraints}
\end{equation}
The factor \(1/2\) multiplying \(P^2\) in the action belongs to the
normalization of the Lagrange multiplier \(e\); it is not part of the
constraint convention.  Likewise, no minus sign is absorbed into the
definition of \(C_2\).  We keep these conventions throughout.
The contraction of the two Virasoro algebras is the classical
\(\BMS_3\), or two-dimensional Galilean conformal algebra
\cite{Bagchi2013}.  Eliminating \(P\) and repackaging \(e,u\) into a
vector density \(V^a\) gives the ILST action
\begin{equation}
 S_{\mathrm{ILST}}
 =\frac{1}{2\pi}\int\dd^2\sigma\,
 V^aV^b\,\partial_aX\cdot\partial_bX,
\label{eq:ILST-action}
\end{equation}
whose degenerate inverse metric is \(V^aV^b\).  The worldsheet has a
preferred null direction and is therefore Carrollian rather than
Lorentzian \cite{Isberg1994,BagchiReview2026}.

The limit in eq.~\eqref{eq:null-phase-space} correctly displays the
traditional two-constraint theory, but it does not by itself decide
whether all redundancies of the degenerate configuration variables have
been divided out.  Degeneracy permits transformations that would be
absent for an invertible worldsheet metric.  In particular, a local
rescaling of the embedding along with an inverse rescaling of the
Carroll vector can preserve the volume density and the action once a
Weyl connection is introduced.  This is the additional
Carroll-Weyl gauge orbit identified in
\cite{SheikhJabbari2026a,SheikhJabbari2026b}.

\subsection*{Carroll-Weyl completion and the third constraint}

Introduce a worldsheet one-form \(W_a\) and the covariant derivative
\begin{equation}
 D_aX^\mu=\partial_aX^\mu+W_aX^\mu .
\label{eq:CW-derivative-early}
\end{equation}
The transformations
\begin{equation}
 X^\mu\longmapsto\e^\chi X^\mu,\qquad
 V^a\longmapsto\e^{-\chi}V^a,\qquad
 W_a\longmapsto W_a-\partial_a\chi
\label{eq:finite-CW}
\end{equation}
make \(V^aD_aX^\mu\) invariant.  The gauged action is therefore
\begin{equation}
 S_{\CW}
 =\frac{1}{2\pi}\int\dd^2\sigma\,
 (V^aD_aX)\cdot(V^bD_bX).
\label{eq:CW-action-early}
\end{equation}
Varying \(W_a\) supplies the new constraint \(P\cdot X=0\).  It is not
a gauge choice imposed on the old system: it is the moment map for an
additional local symmetry.  Consequently its ghost cannot be omitted
without changing the theory being quantized.

This distinction can also be seen directly in phase space.  With the
constraints normalized exactly as in our preceding work \cite{Duary:2026bcs},
\begin{equation}
 C_1[f]=\int\dd\sigma\,fP^2,\qquad
 C_2[g]=\int\dd\sigma\,gP\cdot X',\qquad
 C_3[h]=\int\dd\sigma\,hP\cdot X,
\label{eq:smeared-MLSa}
\end{equation}
the canonical bracket
\(\{X^\mu(\sigma),P_\nu(\sigma')\}
=2\pi\delta^\mu{}_\nu\delta(\sigma-\sigma')\)
gives, up to the harmless common \(2\pi\) convention,
\begin{align}
 \delta_{C_1}X^\mu&=2fP^\mu,&
 \delta_{C_1}P_\mu&=0,
\nonumber\\
 \delta_{C_2}X^\mu&=gX'^\mu,&
 \delta_{C_2}P_\mu&=(gP_\mu)',
\nonumber\\
 \delta_{C_3}X^\mu&=hX^\mu,&
 \delta_{C_3}P_\mu&=-hP_\mu.
\label{eq:constraint-transformations}
\end{align}
The factor of two in the first line and the signs in the second are
direct consequences of the unrescaled definitions.  The last line
makes the semidirect product manifest: \(C_1\) has Carroll-Weyl
charge two, while \(C_2\) transports all fields along the spatial
circle.

The three transformations are independent on a generic point of the
constraint surface.  For each value of \(\sigma\), the \(2D\)-dimensional
canonical phase space is reduced by two dimensions per first-class
constraint.  The local physical phase-space dimension is therefore
\begin{equation}
 2D-2\times3=2(D-3),
\label{eq:dof-count}
\end{equation}
or \(D-3\) configuration-space degrees of freedom.  The old
two-constraint complex instead gives \(D-2\).  The discrepancy is
classical and appears before questions of normal ordering or vacuum
choice arise \cite{SheikhJabbari2026a}.

\subsection*{Quantum consequences of the third constraint}

For a first-class system with constraints \(G_A\), the minimal BFV
charge begins as
\begin{equation}
 Q_{\mathrm{BFV}}
 =c^AG_A-\frac12c^Ac^Bf_{BA}{}^C b_C+\cdots .
\label{eq:BFV-general}
\end{equation}
Thus enlarging \((C_2,C_1)\) to \((C_2,C_1,C_3)\) has three
inseparable effects:
\begin{enumerate}[leftmargin=2em]
\item a fermionic ghost-antighost pair \((s,r)\) is added for the
new bosonic constraint \(C_3\);
\item the nonzero \(C_3\)-\(C_1\) structure constant produces a cubic
interaction \(2s\widetilde b\widetilde c\) in the BRST current;
\item the quantum algebra admits mixed \(LS\) and affine \(SS\)
cocycles in addition to the Virasoro cocycle.
\end{enumerate}
The third point is the decisive one.  A neutral extra \(bc\) pair would
only modify a Virasoro central charge.  The new gauge generator is noncentral in the constraint algebra.
The full ghost currents must therefore realize the same semidirect
product, and BRST nilpotency probes three independent anomaly
coefficients. Table~\ref{tab:complex-comparison} compares the three
string models and shows how gauging the additional constraint \(C_3\)
introduces the ghost pair \((r,s)\) and reduces the number of local
physical configuration modes from \(D-2\) to \(D-3\).

\begin{table}[H]
    \centering
    \renewcommand{\arraystretch}{1.25}
    \setlength{\tabcolsep}{5pt}
    \begin{tabularx}{\linewidth}{%
        |>{\raggedright\arraybackslash}p{0.24\linewidth}%
        |>{\raggedright\arraybackslash}p{0.20\linewidth}%
        |>{\raggedright\arraybackslash}X%
        |>{\centering\arraybackslash}p{0.20\linewidth}|}
        \hline
        \text{String model}
        & \text{Constraints}
        & \text{Minimal ghosts}
        & \text{Physical d.o.f.} \\
        \hline
        Ordinary tensile string
        & Two Virasoro constraints
        & Two chiral \(bc\) systems
        & \(D-2\) \\
        \hline
        Traditional ILST null string
        & \(C_2, C_1\)
        & Two BMS \(bc\)-ghost pairs
        & \(D-2\) \\
        \hline
        Carroll-Weyl gauged null string
        & \(C_2, C_1, C_3\)
        & \((b,c)\), \((\widetilde b,\widetilde c)\), \((r,s)\)
        & \(D-3\) \\
        \hline
    \end{tabularx}
    \caption{First-class constraints, minimal BRST ghost systems, and
             local physical configuration-mode counts in \(D\) target-space
             dimensions.}
    \label{tab:complex-comparison}
\end{table}

The order of operations is therefore important.  Starting from the
two-constraint quantum theory and later imposing \(P\cdot X=0\) on its
states does not reproduce the determinant, ghost interactions, or
central extensions of the three-constraint path integral.  Conversely,
solving all three constraints classically and quantizing only the
reduced variables removes the unreduced BRST current algebra entirely.
These procedures can agree only after a nontrivial equivalence of
measures and operator orderings is demonstrated.

\subsection{Gauge-fixed action and the complete Faddeev-Popov complex}
\label{sec:review}

We now pass from the geometric origin of the extra symmetry to the
precise quantum system it defines.  The action, the three-row
Faddeev-Popov operator, and the equal-time oscillator algebra are not
independent ingredients: together they determine which fields enter the
BRST complex and how the semidirect product is represented.

\subsubsection*{Carroll-Weyl action and constraints}

Let \(X^\mu(\tau,\sigma)\), \(\mu=0,\ldots,D-1\), embed a closed
worldsheet cylinder into flat target space.  The Carroll-Weyl covariant
action is
\begin{equation}
 S_{\CW}[X,V,W]
 =\frac{1}{2\pi}\int\dd^2\sigma\,
 V^aV^b D_aX\cdot D_bX,
 \qquad
 D_aX^\mu=\partial_aX^\mu+W_aX^\mu .
\label{eq:CW-action}
\end{equation}
In the normalization used here, an infinitesimal Carroll-Weyl
transformation acts as
\begin{equation}
 \delta_\chi X^\mu=\chi X^\mu,\qquad
 \delta_\chi V^a=-\chi V^a,\qquad
 \delta_\chi W_a=-\partial_a\chi .
\label{eq:CW-transform}
\end{equation}
A rescaling of \(\chi\) changes all three equations by a common
convention and does not change the physical algebra below.  Under
diffeomorphisms, \(V^a\) is a vector density of weight \(1/2\).

It is useful to define
\begin{equation}
 P^\mu\equiv V^aD_aX^\mu .
\label{eq:P-def}
\end{equation}
The equations obtained by varying \(V^a\) and \(W_a\) are constraints
\begin{equation}
 P\cdot D_aX=0,\qquad P\cdot X=0.
\label{eq:cov-constraints}
\end{equation}
After using the second equation in the first, the independent
constraints can be written as
\begin{equation}
 C_1=P^2,\qquad
 C_2=P\cdot X',\qquad
 C_3=P\cdot X .
\label{eq:three-constraints}
\end{equation}
In the temporal and Carroll-Weyl gauges,
\begin{equation}
 V^a=(1,0),\qquad W\equiv V^aW_a=0,
\label{eq:gauge-slice}
\end{equation}
one has \(P^\mu=\dot X^\mu\) and the matter action becomes
\begin{equation}
 S_X=\frac{1}{2\pi}\int\dd\tau\,\dd\sigma\,\dot X^2 .
\label{eq:matter-gauge-action}
\end{equation}

The smeared constraints obey
\begin{align}
 \{C_1[f],C_1[g]\}&=0,
 &
 \{C_1[f],C_2[g]\}&=C_1[fg'-f'g],
\nonumber\\
 \{C_1[f],C_3[g]\}&=-2C_1[fg],
 &
 \{C_2[f],C_2[g]\}&=C_2[fg'-f'g],
\nonumber\\
 \{C_2[f],C_3[g]\}&=C_3[fg'],
 &
 \{C_3[f],C_3[g]\}&=0 .
\label{eq:smeared-algebra}
\end{align}
For operator products it is convenient to pass to the standard
Weyl-BMS basis
\begin{equation}
 M=\frac12C_1,\qquad L=-C_2,\qquad S=C_3.
\label{eq:constraint-to-LMS}
\end{equation}
This invertible dictionary is derived from the multiplier
normalization and the conventional active sign of the Witt generator;
it does not alter the constraints \(C_i=0\).  With a conventional
Fourier normalization, the corresponding modes satisfy
\begin{align}
 [L_m,L_n]&=(m-n)L_{m+n},
 &
 [L_m,M_n]&=(m-n)M_{m+n},
\nonumber\\
 [L_m,S_n]&=-nS_{m+n},
 &
 [S_m,M_n]&=2M_{m+n},
\nonumber\\
 [M_m,M_n]&=0,
 &
 [S_m,S_n]&=0.
\label{eq:centerless-LMS}
\end{align}
Thus \(M\) has conformal weight two, \(S\) has conformal weight one, and
\(S\) acts on \(M\) with physical charge two.  The \(L,M\) generators
alone form \(\BMS_3\).  The full algebra is the \(\lambda=-1\)
Weyl-BMS algebra \cite{Batlle2024}, with the dilatation generator normalized as
\begin{equation}
 S=2\mathcal{D}.
\label{eq:S-to-D}
\end{equation}
This factor of two will be responsible for factors \(2\) and \(4\) in
the mixed and affine anomaly coefficients.

The ghost pairing fixes the inverse change of basis,
\begin{equation}
 \gamma_1=\frac12\widetilde c,\qquad
 \gamma_2=-c,\qquad
 \gamma_3=s;\qquad
 \beta^1=2\widetilde b,\qquad
 \beta^2=-b,\qquad
 \beta^3=r .
\label{eq:constraint-ghost-dictionary}
\end{equation}
It obeys
\(\gamma_1C_1+\gamma_2C_2+\gamma_3C_3
=\widetilde cM+cL+sS\) and
\(\{\beta^i,\gamma_j\}=\delta^i{}_j\).  We will give all central
coefficients both in this physical constraint basis and in the
standard \((L,M,S)\) basis.

\subsubsection*{Three-row Faddeev-Popov operator}

The gauge functions are
\begin{equation}
 G_0=V^0-1,\qquad G_1=V^1,\qquad G_s=V^aW_a .
\label{eq:gauge-functions}
\end{equation}
If the gauge parameters are ordered as
\(\eta^A=(\epsilon^0,\epsilon^1,\lambda)\), their variations on
eq.~\eqref{eq:gauge-slice} are
\begin{equation}
 \delta G_0=-\half\partial_\tau\epsilon^0
             +\half\partial_\sigma\epsilon^1-\lambda,\qquad
 \delta G_1=-\partial_\tau\epsilon^1,\qquad
 \delta G_s=-\partial_\tau\lambda .
\label{eq:gauge-variations}
\end{equation}
Consequently,
\begin{equation}
 \mathcal{M}_{\CW}=
 \begin{pmatrix}
 -\frac12\partial_\tau & \frac12\partial_\sigma & -1\\
 0 & -\partial_\tau & 0\\
 0 & 0 & -\partial_\tau
 \end{pmatrix}.
\label{eq:FP-operator}
\end{equation}
The upper-left block is the old BMS operator \cite{Chen2023}.  The entry
\((\mathcal{M}_{\CW})_{0s}=-1\) is crucial even though the formal
determinant is triangular: it says that a Carroll-Weyl transformation
moves the gauge condition \(V^0=1\).

Exponentiating \(\det\mathcal{M}_{\CW}\) with ghosts
\((c^0,c^1,s)\) and antighosts \((2b_0,2b_1,b_s)\) gives
\begin{equation}
 S_{bcs}=\frac{\ii}{2\pi}\int\dd^2\sigma\,
 \left(
 c^0\partial_\tau b_0-c^1\partial_\sigma b_0
 +2c^1\partial_\tau b_1+s\partial_\tau b_s-2s b_0
 \right).
\label{eq:bcs-action}
\end{equation}
The equations of motion are
\begin{align}
 \partial_\tau c^1&=0,&
 \partial_\tau s&=0,&
 \partial_\tau c^0-\partial_\sigma c^1+2s&=0,
\nonumber\\
 \partial_\tau b_0&=0,&
 \partial_\tau b_1-\half\partial_\sigma b_0&=0,&
 \partial_\tau b_s-2b_0&=0.
\label{eq:ghost-eom}
\end{align}
The mixing terms in the last equations are the Lagrangian counterpart
of \([S,M]=2M\).

\subsubsection*{Equal-time flipped-vacuum realization}

At a reference slice, which we choose as \(\tau=0\), the independent
oscillators form three canonical pairs
\begin{equation}
 \{b_m,c_n\}=\delta_{m+n,0},\qquad
 \{\widetilde b_m,\widetilde c_n\}=\delta_{m+n,0},\qquad
 \{r_m,s_n\}=\delta_{m+n,0}.
\label{eq:ghost-oscillators}
\end{equation}
Their detailed relation to the cylinder fields is given in
\cref{app:cylinder}.  Introduce the formal Laurent variable
\(z=\e^{\ii\sigma}\).  Since the standard-basis currents \(L,M,S\),
equivalently the physical constraints \(C_2,C_1,C_3\), have conformal weights \(2,2,1\), BRST weight counting fixes
\begin{equation}
 (b,c):(2,-1),\qquad
 (\widetilde b,\widetilde c):(2,-1),\qquad
 (r,s):(1,0).
\label{eq:ghost-weights}
\end{equation}
After choosing the highest-weight splitting, the singular products are
\begin{equation}
 b(z)c(w)\sim\frac{1}{z-w},\qquad
 \widetilde b(z)\widetilde c(w)\sim\frac{1}{z-w},\qquad
 r(z)s(w)\sim\frac{1}{z-w},
\label{eq:ghost-basic-ope}
\end{equation}
with all cross-species products regular.

The matter symplectic form similarly gives the standard bosonic
first-order products
\begin{equation}
 P_\mu(z)X^\nu(w)\sim-\frac{\delta_\mu{}^\nu}{z-w},
 \qquad
 X^\nu(z)P_\mu(w)\sim\frac{\delta_\mu{}^\nu}{z-w}.
\label{eq:matter-basic-ope}
\end{equation}
The relative sign is the radially ordered form of the canonical
commutator.  Differentiating,
\begin{equation}
 \partial X^\mu(z)P_\nu(w)\sim
 -\frac{\delta^\mu{}_\nu}{(z-w)^2},
 \qquad
 P_\nu(z)\partial X^\mu(w)\sim
 -\frac{\delta^\mu{}_\nu}{(z-w)^2}.
\label{eq:matter-derived-ope}
\end{equation}
Eqs.~\eqref{eq:ghost-basic-ope} and
\eqref{eq:matter-basic-ope} are the raw data for the OPE computation.

\subsection{Equal-time OPE prescription}
\label{sec:ope-framework}

The remaining conceptual step is to turn the canonical cylinder data
into an operator algebra without treating the Carrollian worldsheet as a
relativistic CFT.  This section fixes that dictionary and the
normal-ordering prescription before any anomaly is evaluated.

\subsubsection*{The formal Laurent variable on a Carrollian slice}

The familiar OPE of a relativistic CFT is supported by complex
analyticity and by a separation into left- and right-movers.  Neither
statement holds automatically on a degenerate worldsheet.  Our use of
\(z\) has a narrower and fully canonical meaning, following the
fixed-time construction used for the BMS algebra in
\cite{FigueroaHaveObers2025}.  At fixed \(\tau\),
the spatial circle carries the Witt action generated by \(L_n\).
Fourier modes may therefore be collected into Laurent fields of
definite conformal weight.  Radial ordering in the auxiliary \(z\)-plane is
then a convenient encoding of the ordered mode algebra.

\paragraph{Equal-time slice and Carrollian evolution.}
The explicit contractions below are evaluated on the reference slice
$\tau=0$, with $z=e^{\ii\sigma}$ used to organize the spatial Fourier
modes into Laurent fields. This restriction does not discard the
Carrollian-time dependence of the theory. At nonzero $\tau$, one may
introduce the second Carrollian-plane coordinate
$u=\ii\tau e^{\ii\sigma}=\ii\tau z$ and recover the full two-coordinate
fields from their equations of motion. This point is particularly
important for the ghost sector: although the mixing term $-2sb^{0}$
does not modify the equal-time canonical anticommutators, it couples the
equations of motion and therefore changes the time evolution of the
component fields and composite currents. The equal-time OPEs must
accordingly be understood as action-derived initial data whose
Carrollian evolution gives the full-plane current algebra discussed in \cref{app:two-coordinate}.

For the matter pair,
\begin{equation}
 X^\mu(z)=\sum_{n\in\mathbb Z}X^\mu_nz^{-n},
 \qquad
 P_\mu(z)=\sum_{n\in\mathbb Z}P_{\mu,n}z^{-n-1},
\label{eq:XP-Laurent-main}
\end{equation}
and the equal-time symplectic form gives
\begin{equation}
 [X^\mu_m,P_{\nu,n}]
 =\delta^\mu{}_\nu\delta_{m+n,0}.
\label{eq:XP-mode-bracket}
\end{equation}
A highest-weight splitting assigns positive-frequency modes to
annihilation operators, with a separate prescription for zero modes.
Summing the resulting geometric series for \(|z|>|w|\) produces
eq.~\eqref{eq:matter-basic-ope}.  Exchanging the radial order reverses the
canonical commutator and explains the sign difference between
\(P(z)X(w)\) and \(X(z)P(w)\).

The same construction applies to the three fermionic ghost pairs.  A
field of weight \(h\) is expanded as
\(\Phi(z)=\sum_n\Phi_nz^{-n-h}\), and the anticommutators in
eq.~\eqref{eq:ghost-oscillators} sum to the three poles in
eq.~\eqref{eq:ghost-basic-ope}. The full dependence on Carrollian time is
restored by the evolution equations in
\cref{app:two-coordinate}.

\subsubsection*{Normal ordering and Wick signs}

Normal ordering is defined by moving annihilation modes to the right,
with a minus sign whenever two fermionic operators are exchanged.
For two composite operators, Wick's theorem reads schematically
\begin{equation}
\begin{aligned}
 \normal{A_1\cdots A_p}(z)\,
 \normal{B_1\cdots B_q}(w)
 &=
 \sum_{\text{cross-contractions}}
 (\pm)\,\big(A_i(z)B_j(w)\big)_{\mathrm{sing}}
 \\
 &\hspace{4.2cm}\times
 \normal{\text{uncontracted fields}},
\end{aligned}
\label{eq:wick-schematic}
\end{equation}
where contractions occur only between the two normal-ordered blocks.
The signs in fermionic double contractions are especially important:
the central charge of a \(bc\) system is not obtained by multiplying
two bosonic contractions.

The differentiated matter contractions used repeatedly below are
\begin{align}
 P_\mu(z)\partial^kX^\nu(w)
 &\sim
 -\frac{k!\,\delta_\mu{}^\nu}{(z-w)^{k+1}},
\nonumber\\
 \partial^kX^\nu(z)P_\mu(w)
 &\sim
 \frac{(-1)^kk!\,\delta_\mu{}^\nu}{(z-w)^{k+1}},
\label{eq:derived-contractions-general}
\end{align}
with the \(k=1\) signs agreeing with
eq.~\eqref{eq:matter-derived-ope}.  For a fermionic pair
\(B(z)C(w)\sim(z-w)^{-1}\), derivatives follow by differentiating the
radially ordered kernel before taking the coincident expansion.  We use
point splitting,
\begin{equation}
 \normal{AB}(w)
 =\lim_{z\to w}
 \left[A(z)B(w)-\big(A(z)B(w)\big)_{\mathrm{sing}}\right],
\label{eq:point-splitting}
\end{equation}
so all finite composite operators are defined in the same
regularization.

The Taylor expansion of an uncontracted field at \(z\),
\begin{equation}
 \Phi(z)=\sum_{k\geq0}\frac{(z-w)^k}{k!}\,
 \partial^k\Phi(w),
\label{eq:Taylor-OPE}
\end{equation}
is what turns single contractions into the current and derivative terms
of a Ward identity.  Double contractions contain no uncontracted fields
and determine the central poles.  This separation makes the calculation
auditable: closure of the single contractions checks the classical
algebra, while the double contractions measure its quantum anomaly.

\subsubsection*{Current-algebra conventions}

We normalize the stress tensor so that a field \(\Phi_h\) of conformal weight
\(h\) obeys
\begin{equation}
 T(z)\Phi_h(w)\sim
 \frac{h\Phi_h(w)}{(z-w)^2}
 +
 \frac{\partial\Phi_h(w)}{z-w}.
\label{eq:primary-ward}
\end{equation}
The three possible central OPEs are written
\begin{align}
 T(z)T(w)&\sim\frac{\cL/2}{(z-w)^4}+\cdots,
\nonumber\\
 T(z)S(w)&\sim\frac{\cS}{(z-w)^3}+\cdots,
\nonumber\\
 S(z)S(w)&\sim\frac{\cSS}{(z-w)^2}+\cdots.
\label{eq:central-ope-conventions}
\end{align}
The orders of the poles are fixed by the weights \(2,2,1\) of
\(T,M,S\).  A fourth-order \(TM\) pole would be allowed by weights
alone, but the Jacobi identity with \(S_0\) removes it from the full
algebra.  No central pole can appear in \(MM\) or \(SM\) in the
present realization \cite{Batlle2024}.

The mixed coefficient \(\cS\) is sometimes called a background charge
or a Virasoro-current anomaly, whereas \(\cSS\) is the level of the
abelian current \(S\).  They are cohomologically distinct from
\(\cL\).  This is the first major difference from the tensile calculation, where cancellation of one Virasoro central
charge suffices.  It also explains why checking only the total
\(T(z)T(w)\) OPE cannot establish nilpotency here.

\subsubsection*{Representation and regularization data}

Canonical brackets determine commutators but not a vacuum expectation
value.  The contraction kernel depends on which half of the modes
annihilates the vacuum and on how divergent sums are regulated.
Historical null-string quantizations already exhibited this
sensitivity: operator orderings leading to a critical dimension can
coexist with admissible orderings that do not select one
\cite{Lizzi1986,Bozhilov1997}.  More recently, the conventional
two-constraint null string was shown to possess inequivalent oscillator,
induced, and flipped vacua with different critical-dimension statements
\cite{BagchiVacua2021}.

Our prescription is deliberately fixed before any contraction is
performed:
\begin{enumerate}[leftmargin=2em]
\item use the canonical equal-time Laurent fields;
\item choose the highest-weight splitting;
\item define composites by the point-splitting subtraction
eq.~\eqref{eq:point-splitting};
\item use the same ordering for matter, all three ghost pairs, and the
BRST current.
\end{enumerate}
 The result below is therefore a statement about a definite quantum
 representation.  

\section{Matter and ghost currents, operator products, and anomaly coefficients}
\label{sec:ope}

This section is the computational core of the paper.  Single
contractions must reproduce the centerless Weyl-BMS brackets; double
contractions then isolate the three quantum obstructions.  Organizing
the calculation in this order cleanly separates a closure check from an
anomaly calculation.  We write
\begin{equation}
 u\equiv z-w
\label{eq:u-definition}
\end{equation}
throughout this section.

\subsection{Matter currents and their anomaly coefficients}
\label{subsec:matter-ope}

We first determine the matter currents directly from the Carroll-Weyl gauged action. Their form is fixed by the response of the action to variations of the auxiliary worldsheet fields and therefore follows from the underlying gauge structure.  With
\(P=V^aD_aX\), a general variation of the auxiliary fields gives
\begin{equation}
 \delta_{V,W}S_{\CW}
 =\frac1\pi\int\dd^2\sigma\,
 \left[
  P\cdot D_aX\,\delta V^a
  +(P\cdot X)V^a\delta W_a
 \right].
\label{eq:first-principles-response}
\end{equation}
On the gauge slice \(V^a=(1,0)\), \(V^aW_a=0\), the three independent
responses are \(P^2\), \(P\cdot X'\), and \(P\cdot X\).  Thus
eq.~\eqref{eq:first-principles-response} derives,
\begin{equation}
 C_1=P^2,\qquad C_2=P\cdot X',\qquad C_3=P\cdot X .
\label{eq:first-principles-C}
\end{equation}
The same statement is especially transparent in Hamiltonian form
\begin{equation}
 S_{\rm ph}=\frac1{2\pi}\int\dd\tau\,\dd\sigma\,
 \left[
 P\cdot\dot X-\frac e2C_1-uC_2-aC_3
 \right].
\label{eq:first-principles-phase-space}
\end{equation}
The multiplier derivatives give
\(\delta S_{\rm ph}/\delta e=-C_1/(4\pi)\),
\(\delta S_{\rm ph}/\delta u=-C_2/(2\pi)\), and
\(\delta S_{\rm ph}/\delta a=-C_3/(2\pi)\).  The factor \(1/2\) in
the first relation is therefore a multiplier convention, not a change
from \(C_1=P^2\).

The canonical one-form in
eq.~\eqref{eq:first-principles-phase-space} gives
\(\{X^\mu(\sigma),P_\nu(\sigma')\}
=2\pi\delta^\mu{}_\nu\delta(\sigma-\sigma')\).  The exact smeared
constraint generators are
\begin{equation}
 C_1[f]=\int\dd\sigma\,fP^2,\qquad
 C_2[g]=\int\dd\sigma\,g\,P\cdot X',\qquad
 C_3[h]=\int\dd\sigma\,h\,X\cdot P .
\label{eq:action-moment-maps}
\end{equation}
For example,
\begin{align}
 \{X^\mu,C_2[g]\}&=g\,\partial_\sigma X^\mu,
 &
 \{P_\mu,C_2[g]\}&=\partial_\sigma(g P_\mu),
\nonumber\\
 \{X^\mu,C_3[h]\}&=h X^\mu,
 &
 \{P_\mu,C_3[h]\}&=-h P_\mu ,
\label{eq:moment-map-check}
\end{align}
which are precisely the transformations inherited from the
Carroll-Weyl gauged action.  Hence \(C_2\) is the spatial
reparametrization response -- the Carrollian analogue of the momentum
component of the stress tensor -- and \(C_1\) is the null Hamiltonian
response.  No inverse worldsheet metric is required on the degenerate
surface.

After the cylinder-to-plane tensor-density factors are absorbed into
the Laurent fields, point splitting promotes the physical constraints
to
\begin{equation}
 \mathcal C_1^X=\normal{P_\mu P^\mu},\qquad
 \mathcal C_2^X=\normal{P_\mu\partial X^\mu},\qquad
 \mathcal C_3^X=\normal{X^\mu P_\mu}.
\label{eq:physical-matter-currents}
\end{equation}
The standard active Witt convention and
eq.~\eqref{eq:constraint-to-LMS} then give the OPE-adapted currents
\begin{equation}
 T^X=-\normal{P_\mu\partial X^\mu},\qquad
 M^X=\half\normal{P_\mu P^\mu},\qquad
 S^X=\normal{X^\mu P_\mu}.
\label{eq:matter-currents}
\end{equation}
The sign and factor \(1/2\) in this last display belong only to the
derived \((L,M,S)\) basis.  The physical constraints remain exactly
those of eq.~\eqref{eq:first-principles-C}.  The OPE calculation below
is a quantum test of these action-derived currents, not their
definition.

\subsubsection*{Elementary Ward identities}

Before multiplying two composite currents, it is useful to check their
action on the elementary fields.  A single contraction gives
\begin{align}
 T^X(z)X^\mu(w)
 &\sim\frac{\partial X^\mu(w)}{z-w},
 &
 T^X(z)P_\mu(w)
 &\sim\frac{P_\mu(w)}{(z-w)^2}
       +\frac{\partial P_\mu(w)}{z-w},
\label{eq:T-on-XP}\\
 S^X(z)X^\mu(w)
 &\sim-\frac{X^\mu(w)}{z-w},
 &
 S^X(z)P_\mu(w)
 &\sim\frac{P_\mu(w)}{z-w}.
\label{eq:S-on-XP}
\end{align}
Thus \(X\) and \(P\) have conformal weights \(0\) and \(1\), and
\(S\)-charges \(-1\) and \(+1\), respectively.  The charge convention
is the adjoint convention in which \([S,M]=2M\); the active canonical
variation generated with \(S\) in the second Poisson-bracket slot has
the opposite sign.

Eqs.~\eqref{eq:T-on-XP} and \eqref{eq:S-on-XP} already predict all
non-central terms in the matter algebra.  They imply that \(M^X\) has
weight two and \(S\)-charge two, while \(S^X\) has weight one.  What
they do not determine are the central poles, which require the double
contractions below.

\subsubsection*{\(T^X T^X\).}

Using the differentiated elementary products
eq.~\eqref{eq:matter-derived-ope}, the double contraction is
\begin{align}
 T^X(z)T^X(w)\big|_{\mathrm{double}}
 &=
 \big(P_\mu(z)\partial X^\nu(w)\big)_{\mathrm{sing}}
 \big(\partial X^\mu(z)P_\nu(w)\big)_{\mathrm{sing}}
\nonumber\\
 &=\left(-\frac{\delta_\mu{}^\nu}{u^2}\right)
   \left(-\frac{\delta^\mu{}_\nu}{u^2}\right)
 =\frac{D}{u^4}.
\label{eq:TXTX-double}
\end{align}
The two single contractions retain the uncontracted normal-ordered
fields
\begin{equation}
 T^X(z)T^X(w)\big|_{\mathrm{single}}
 =-\frac{\normal{\partial X^\mu(z)P_\mu(w)}}{u^2}
  -\frac{\normal{P_\mu(z)\partial X^\mu(w)}}{u^2}.
\label{eq:TXTX-single-raw}
\end{equation}
Taylor expansion at \(w\) gives
\begin{align}
 \normal{\partial X^\mu(z)P_\mu(w)}
 &=\normal{\partial X^\mu P_\mu}(w)
   +u\normal{\partial^2X^\mu P_\mu}(w)+O(u^2),
\nonumber\\
 \normal{P_\mu(z)\partial X^\mu(w)}
 &=\normal{P_\mu\partial X^\mu}(w)
   +u\normal{(\partial P_\mu)\partial X^\mu}(w)+O(u^2).
\label{eq:TXTX-Taylor}
\end{align}
Because the matter fields are bosonic,
\(\normal{\partial X\cdot P}=\normal{P\cdot\partial X}=-T^X\), while
\begin{equation}
 \partial T^X
 =-\normal{(\partial P)\cdot\partial X}
  -\normal{P\cdot\partial^2X}.
\label{eq:derivative-TX}
\end{equation}
Substitution into eq.~\eqref{eq:TXTX-single-raw} therefore yields
\begin{equation}
 T^X(z)T^X(w)\big|_{\mathrm{single}}
 =\frac{2T^X(w)}{u^2}+\frac{\partial T^X(w)}{u}.
\label{eq:TXTX-single-final}
\end{equation}
Combining the double and single contractions gives
\begin{equation}
 T^X(z)T^X(w)\sim
 \frac{D}{u^4}
 +\frac{2T^X(w)}{u^2}
 +\frac{\partial T^X(w)}{u}.
\label{eq:TXTX}
\end{equation}
In the convention in which the fourth-order pole is \(\cL/2\),
\begin{equation}
 \cL^X=2D.
\label{eq:cLL-matter}
\end{equation}
This is the first-order analogue of the familiar contribution \(D\)
from \(D\) second-order scalars.

The factor of two has a simple origin.  Each target-space index gives
one bosonic first-order \((P,X)\) system of weights \((1,0)\), whose
stress tensor has \(c=2\).  This should not be confused with two
propagating worldsheet scalars: \(X\) and \(P\) are conjugate variables,
and the null equations of motion are first order in Carrollian time.

\subsubsection*{\(T^X S^X\) and \(S^X S^X\).}

For the mixed OPE, the two simultaneous contractions are
\begin{align}
 \left[-P_\mu(z)\partial X^\mu(z)\right]
 \left[X^\nu(w)P_\nu(w)\right]
 \Big|_{\mathrm{double}}
 &=-\left(-\frac{\delta_\mu{}^\nu}{u}\right)
       \left(-\frac{\delta^\mu{}_\nu}{u^2}\right)
\nonumber\\
 &=-\frac{D}{u^3}.
\label{eq:TXSX-double}
\end{align}
The single contractions are equally important because they fix the
non-central transformation law,
\begin{equation}
 T^X(z)S^X(w)\big|_{\mathrm{single}}
 =\frac{\normal{\partial X^\mu(z)P_\mu(w)}}{u}
  +\frac{\normal{P_\mu(z)X^\mu(w)}}{u^2}.
\label{eq:TXSX-single-raw}
\end{equation}
At the orders that remain singular,
\begin{align}
 \normal{\partial X^\mu(z)P_\mu(w)}
 &=-T^X(w)+O(u),
\nonumber\\
 \normal{P_\mu(z)X^\mu(w)}
 &=S^X(w)+u\normal{X^\mu\partial P_\mu}(w)+O(u^2).
\label{eq:TXSX-Taylor}
\end{align}
Since
\begin{equation}
 \partial S^X
 =\normal{\partial X\cdot P}+\normal{X\cdot\partial P}
 =-T^X+\normal{X\cdot\partial P},
\label{eq:derivative-SX}
\end{equation}
eq.~\eqref{eq:TXSX-single-raw} becomes
\(S^X/u^2+\partial S^X/u\).  Adding the double contraction gives
\begin{equation}
 T^X(z)S^X(w)\sim
 -\frac{D}{u^3}
 +\frac{S^X(w)}{u^2}
 +\frac{\partial S^X(w)}{u}.
\label{eq:TXSX}
\end{equation}
The \(S^X S^X\) double contraction has the opposite radial-ordering
sign between \(XP\) and \(PX\),
\begin{equation}
 S^X(z)S^X(w)\big|_{\mathrm{double}}
 =
 \left(\frac{\delta^\mu{}_\nu}{u}\right)
 \left(-\frac{\delta_\mu{}^\nu}{u}\right)
 =-\frac{D}{u^2}.
\label{eq:SXSX-double}
\end{equation}
The two single contractions are
\begin{equation}
 S^X(z)S^X(w)\big|_{\mathrm{single}}
 =\frac{\normal{P_\mu(z)X^\mu(w)}}{u}
  -\frac{\normal{X^\mu(z)P_\mu(w)}}{u}.
\label{eq:SXSX-single}
\end{equation}
At leading order both normal products equal \(S^X(w)\), so the two
possible simple poles cancel.  Consequently
\begin{equation}
 S^X(z)S^X(w)\sim-\frac{D}{u^2}.
\label{eq:SXSX}
\end{equation}
Hence
\begin{equation}
 \cS^X=-D,\qquad \cSS^X=-D.
\label{eq:matter-scaling-anomalies}
\end{equation}

The two coefficients have different meanings even though they happen
to be numerically equal in this free realization.  The third-order pole
measures the failure of \(S^X\) to be a primary current under \(T^X\);
the second-order pole is the level of the scaling current.  Adding an
independent sector or improving the stress tensor need not preserve
their equality.

\subsubsection*{OPEs involving \(M^X\)}

Because \(P_\mu\) is a weight-one field under \(T^X\), Wick's theorem
immediately gives
\begin{align}
 T^X(z)M^X(w)
 &\sim \frac{2M^X(w)}{(z-w)^2}
       +\frac{\partial M^X(w)}{z-w},
\label{eq:TXMX}\\
 S^X(z)M^X(w)
 &\sim \frac{2M^X(w)}{z-w},
\label{eq:SXMX}\\
 M^X(z)M^X(w)&\sim 0.
\label{eq:MXMX}
\end{align}
There is no \(P\)-\(P\) contraction.  Notice in particular the absence
of an \(LM\) fourth-order pole.

One can see this absence without any detailed expansion.  Every term in
\(M^X\) contains only \(P\)'s.  A double contraction with \(T^X\)
would require contracting both the \(P\) and the \(\partial X\) in
\(T^X\), but \(P(z)P(w)\) is regular.  The same triangularity makes
\(M^X(z)M^X(w)\) regular.  This free-field observation is reinforced
algebraically by the Jacobi identity in \cref{sec:brst}.

Returning now to the physical constraints rather than the
OPE-adapted basis, use
\(\mathcal C_1^X=2M^X\),
\(\mathcal C_2^X=-T^X\), and
\(\mathcal C_3^X=S^X\).  No contractions need to be recomputed: direct
substitution into eqs.~\eqref{eq:TXTX}, \eqref{eq:TXSX},
\eqref{eq:SXSX}, and \eqref{eq:TXMX}--\eqref{eq:MXMX} gives
\begin{align}
 \mathcal C_2^X(z)\mathcal C_2^X(w)
 &\sim \frac{D}{u^4}
 -\frac{2\mathcal C_2^X(w)}{u^2}
 -\frac{\partial\mathcal C_2^X(w)}{u},
\label{eq:C2C2-matter}\\
 \mathcal C_2^X(z)\mathcal C_3^X(w)
 &\sim \frac{D}{u^3}
 -\frac{\mathcal C_3^X(w)}{u^2}
 -\frac{\partial\mathcal C_3^X(w)}{u},
\label{eq:C2C3-matter}\\
 \mathcal C_3^X(z)\mathcal C_3^X(w)
 &\sim-\frac{D}{u^2},
\label{eq:C3C3-matter}\\
 \mathcal C_2^X(z)\mathcal C_1^X(w)
 &\sim-\frac{2\mathcal C_1^X(w)}{u^2}
       -\frac{\partial\mathcal C_1^X(w)}{u},
\label{eq:C2C1-matter}\\
 \mathcal C_3^X(z)\mathcal C_1^X(w)
 &\sim\frac{2\mathcal C_1^X(w)}{u},
\qquad
 \mathcal C_1^X(z)\mathcal C_1^X(w)\sim0 .
\label{eq:C3C1-matter}
\end{align}
The minus signs in the \(C_2\) Ward identities simply record that the
standard Witt generator is \(L=-C_2\).  In particular, the central
coefficients in the physical \((C_2,C_3)\) basis are
\begin{equation}
 (k_{22},k_{23},k_{33})_X=(2D,D,-D),
\label{eq:physical-matter-anomalies}
\end{equation}
where the fourth-, third-, and second-order poles are
\(k_{22}/2\), \(k_{23}\), and \(k_{33}\), respectively.

Collecting the independent matter anomalies,
\begin{equation}
 \big(\cL,\cS,\cSS\big)_X=(2D,-D,-D).
\label{eq:matter-anomaly-vector}
\end{equation}
Equivalently, if one uses the unit-charge Weyl-BMS current
\(\mathcal{D}=S/2\), then
\((c_{L\mathcal{D}},c_{\mathcal{D}\mathcal{D}})
=(-D/2,-D/4)\), in agreement with the first-order free-field
realization of \cite{Batlle2024}.

\subsection{Ghost currents and their anomaly coefficients}

Assign ghost number \(+1\) to \(c,\widetilde c,s\) and \(-1\) to
\(b,\widetilde b,r\).  Their weights, canonical partners, and roles are
collected in \cref{tab:ghost-data}.  The scalar ghost \(s\) is scalar
only under the spatial Witt algebra; it is still Grassmann odd and
carries the same BRST ghost number as the diffeomorphism ghosts.  The
canonical pairings follow directly from the time-derivative terms in
the Faddeev-Popov action eq.~\eqref{eq:bcs-action}.  After the field rescalings recorded in \cref{app:cylinder}, its equal-time
symplectic form is
\begin{equation}
 \Omega_{\gh}
 =\frac{\ii}{2\pi}\int\dd\sigma\,
 \big(\delta b\wedge\delta c
     +\delta\widetilde b\wedge\delta\widetilde c
     +\delta r\wedge\delta s\big),
\label{eq:ghost-symplectic-form}
\end{equation}
which yields the three elementary contractions in
eq.~\eqref{eq:ghost-basic-ope}.  The algebraic coupling \(-2sb_0\)
contains no time derivative.  It changes the Hamiltonian evolution and
the off-diagonal ghost currents, but not this equal-time symplectic
form.

\begin{table}[H]
\centering
\renewcommand{\arraystretch}{1.16}
\begin{tabular}{|c|c|c|c|c|}
\hline
Constraint & Weight & Ghost & Antighost & Ghost weights \\
\hline
\(C_2=P\cdot X'\;(L=-C_2)\) & \(2\) & \(c\) & \(b\)
& \((-1,2)\) \\
\(C_1=P^2\;(M=C_1/2)\) & \(2\) & \(\widetilde c\) & \(\widetilde b\)
& \((-1,2)\) \\
\(C_3=P\cdot X\;(S=C_3)\) & \(1\) & \(s\) & \(r\)
& \((0,1)\) \\
\hline
\end{tabular}
\caption{The minimal ghost pairs. Each ghost has weight
\(1-h_G\) and its antighost has weight \(h_G\), so
\(\normal{c^GG}\) has weight one in the BRST current.}
\label{tab:ghost-data}
\end{table}

We next derive the ghost stress tensors.  For a
fermionic first-order pair \((B,C)\) with
\(B(z)C(w)\sim C(z)B(w)\sim u^{-1}\), take the most general local
bilinear of weight two,
\begin{equation}
 T_h=a\normal{B\partial C}+d\normal{(\partial B)C}.
\label{eq:generic-ghost-T-ansatz}
\end{equation}
The constraint associated with \(B\) has conformal weight \(h\); hence its
ghost has weight \(1-h\).  Requiring the Ward identities
\begin{align}
 T_h(z)B(w)&\sim\frac{hB(w)}{u^2}
                    +\frac{\partial B(w)}{u},
\nonumber\\
 T_h(z)C(w)&\sim\frac{(1-h)C(w)}{u^2}
                    +\frac{\partial C(w)}{u}
\label{eq:generic-ghost-Ward}
\end{align}
fixes \(a=-h\) and \(d=1-h\).  Therefore the Noether stress tensor
of the first-order ghost action is
\begin{equation}
 T_h=-h\normal{B\partial C}
 -(h-1)\normal{(\partial B)C}.
\label{eq:first-order-stress}
\end{equation}
This derivation also shows why the diagonal Witt current is unaffected
by the off-diagonal term in eq.~\eqref{eq:bcs-action}: its coefficients
are fixed by the canonical kinetic term and the transformation weights.

For completeness, the fourth-order pole can be evaluated without
invoking a tabulated \(bc\) central charge.  Define
\begin{equation}
 \mathcal A=\normal{(\partial B)C},\qquad
 \mathcal B=\normal{B\partial C},\qquad
 T_h=(1-h)\mathcal A-h\mathcal B .
\label{eq:ghost-AB}
\end{equation}
Including the Grassmann Wick signs, the double-contraction contributions are
\begin{align}
 \mathcal A(z)\mathcal A(w)\big|_{\rm double}
 &=-\frac1{u^4},
 &
 \mathcal B(z)\mathcal B(w)\big|_{\rm double}
 &=-\frac1{u^4},
\nonumber\\
 \big[\mathcal A(z)\mathcal B(w)
 +\mathcal B(z)\mathcal A(w)\big]_{\rm double}
 &=-\frac4{u^4}.
\label{eq:ghost-double-ledger}
\end{align}
It follows that
\begin{align}
 T_h(z)T_h(w)\big|_{\rm double}
 &=\frac{-6h^2+6h-1}{u^4}
\nonumber\\
 &=\frac{c_{\gh}(h)/2}{u^4},
\label{eq:generic-ghost-fourth-pole}\\
 c_{\gh}(h)
 &=1-3(2h-1)^2
  =-2(6h^2-6h+1).
\label{eq:bc-central-charge}
\end{align}
The single contractions reproduce
\(2T_h/u^2+\partial T_h/u\), as required by
eq.~\eqref{eq:generic-ghost-Ward}.  Thus the result is
\begin{equation}
 T_h(z)T_h(w)\sim
 \frac{c_{\gh}(h)/2}{u^4}
 +\frac{2T_h(w)}{u^2}
 +\frac{\partial T_h(w)}{u}.
\label{eq:generic-ghost-TT}
\end{equation}
In particular, the weight-\((1,0)\) pair contributes
\(-\normal{r\partial s}\) with no
\(\normal{(\partial r)s}\) term.
For the three pairs in eq.~\eqref{eq:ghost-weights},
\begin{equation}
 c_{bc}=-26,\qquad
 c_{\widetilde b\widetilde c}=-26,\qquad
 c_{rs}=-2.
\label{eq:three-ghost-cc}
\end{equation}
Thus
\begin{align}
 T^\gh={}&
 -2\normal{b\partial c}-\normal{(\partial b)c}
 -\normal{r\partial s}
\nonumber\\
 &-2\normal{\widetilde b\partial\widetilde c}
 -\normal{(\partial\widetilde b)\widetilde c},
\label{eq:Tghost}\\
 T^\gh(z)T^\gh(w)\sim{}&
 -\frac{27}{(z-w)^4}
 +\frac{2T^\gh(w)}{(z-w)^2}
 +\frac{\partial T^\gh(w)}{z-w},
\label{eq:TghostTghost}
\end{align}
and
\begin{equation}
 \cL^\gh=-54.
\label{eq:cLL-ghost}
\end{equation}
This is already enough to see that a \(TT\)-only check would select
\(2D-54=0\), or \(D=27\).  The remaining current algebra shows why
that check is incomplete.

\subsection*{Semidirect-product ghost currents}
\label{subsec:ghost-currents}

The ghost currents are fixed by the structure constants, not by treating
the three pairs independently.  In local notation the BRST differential
on the ghosts is
\begin{align}
 \delta_B c&=c\,\partial c,
\label{eq:brst-c}\\
 \delta_B s&=c\,\partial s,
\label{eq:brst-s}\\
 \delta_B\widetilde c
 &=c\,\partial\widetilde c
 +\widetilde c\,\partial c+2s\widetilde c.
\label{eq:brst-ctilde}
\end{align}
The last term is forced by \([S,M]=2M\).  Contracting the BFV current
with the three antighosts, or equivalently mapping the standard
\(\lambda=-1\) Weyl-BMS construction to \(S=2\mathcal D\), yields
\begin{align}
 S^\gh&=
 2\normal{\widetilde b\widetilde c}
 +\partial\normal{c r},
\label{eq:Sghost}\\
 M^\gh&=
 \normal{c\,\partial\widetilde b}
 +2\normal{s\widetilde b}
 +2\normal{(\partial c)\widetilde b}.
\label{eq:Mghost}
\end{align}
The terms \(\partial(cr)\) and \(2s\widetilde b\) are essential.
They are the spatial current-algebra image of the cylinder coupling
\(-2sb_0\) in eq.~\eqref{eq:bcs-action}.

There is a useful conceptual distinction between
\(T^\gh\) and \(S^\gh,M^\gh\).  The stress tensor is the sum of the
three free first-order stress tensors because the Witt generator acts
diagonally on the three canonical pairs.  The other two currents are
not diagonal sums.  They implement the adjoint action of the complete
constraint algebra on its ghosts, so the off-diagonal structure
constant \(f_{SM}{}^M=2\) necessarily mixes the \(S\)-ghost with the
\(M\)-antighost.  Treating the scalar pair as a spectator with its own
number current would give the correct \(-2\) contribution to
\(\cL^\gh\) but the wrong \(LS\), \(SS\), and \(SM\) OPEs.

We now evaluate the singular products.  Define
\begin{equation}
 N_M=\normal{\widetilde b\widetilde c},\qquad
 A=\normal{cr},
\qquad S^\gh=2N_M+\partial A.
\label{eq:NM-A}
\end{equation}
The third-order pole in \(T^\gh S^\gh\) comes from the
weight-\((2,-1)\) \((\widetilde b,\widetilde c)\) pair:
\begin{align}
 &\left[-2\normal{\widetilde b\partial\widetilde c}(z)\right]
   \left[2\normal{\widetilde b\widetilde c}(w)\right]
   \Big|_{\rm double}
 =\frac4{u^3},
\nonumber\\
 &\left[-\normal{(\partial\widetilde b)\widetilde c}(z)\right]
   \left[2\normal{\widetilde b\widetilde c}(w)\right]
   \Big|_{\rm double}
 =\frac2{u^3}.
\label{eq:TghSgh-two-contributions}
\end{align}
Therefore
\begin{equation}
 T^\gh(z)\,2N_M(w)\big|_{\mathrm{double}}
 =\frac{4+2}{u^3}=\frac6{u^3}.
\label{eq:TghSgh-double}
\end{equation}
The \(\partial A\) term completes the non-central transformation law
but creates no field-independent third-order pole: \(A\) involves only
the independent \((b,c)\) and \((r,s)\) pairs, so it cannot doubly
contract with either individual stress-tensor term.  Therefore
\begin{equation}
 T^\gh(z)S^\gh(w)\sim
 \frac{6}{u^3}
 +\frac{S^\gh(w)}{u^2}
 +\frac{\partial S^\gh(w)}{u}.
\label{eq:TghSgh}
\end{equation}

In \(S^\gh(z)S^\gh(w)\), the only double contraction comes from
\(4N_M(z)N_M(w)\),
\begin{equation}
 N_M(z)N_M(w)\big|_{\rm double}
 =\big\langle\widetilde b(z)\widetilde c(w)\big\rangle
  \big\langle\widetilde c(z)\widetilde b(w)\big\rangle
 =\frac1{u^2}.
\label{eq:NMNM-double}
\end{equation}
The mixed terms are regular because \(N_M\) contains only
\((\widetilde b,\widetilde c)\), whereas \(A\) contains only
\((c,r)\); moreover \(\partial A(z)\partial A(w)\) has no double
contraction.  Hence
\begin{equation}
 S^\gh(z)S^\gh(w)\sim\frac{4}{u^2}.
\label{eq:SghSgh}
\end{equation}

The remaining products reproduce the centerless semidirect action,
\begin{align}
 T^\gh(z)M^\gh(w)
 &\sim \frac{2M^\gh(w)}{(z-w)^2}
       +\frac{\partial M^\gh(w)}{z-w},
\label{eq:TghMgh}\\
 S^\gh(z)M^\gh(w)
 &\sim \frac{2M^\gh(w)}{z-w},
\label{eq:SghMgh}\\
 M^\gh(z)M^\gh(w)&\sim0.
\label{eq:MghMgh}
\end{align}
The term-by-term derivation is recorded in \cref{app:ghost}.
The independent ghost anomalies are therefore
\begin{equation}
 \big(\cL,\cS,\cSS\big)_{\gh}=(-54,6,4).
\label{eq:ghost-anomaly-vector}
\end{equation}

The three ghost anomaly coefficients have different origins.  The
Virasoro-type coefficient follows directly from the central charge of a
fermionic first-order system,
\[
c_{\gh}(h)=1-3(2h-1)^2.
\]
The Carroll-Weyl ghost complex contains two weight-\((2,-1)\) systems,
\((b,c)\) and \((\widetilde b,\widetilde c)\), together with one
weight-\((1,0)\) system, \((r,s)\).  Therefore,
\[
c_{LL}^{\gh}
=
2c_{\gh}(2)+c_{\gh}(1)
=
2(-26)+(-2)
=
-54.
\]
The remaining coefficients follow from the ghost Carroll-Weyl current
\[
S^{\gh}=2N_M+\partial A,
\qquad
N_M=:\widetilde b\widetilde c:,
\qquad
A=:cr:.
\]
For a weight-\((h,1-h)\) ghost system, the number current satisfies
\[
T_h(z)N(w)\sim
\frac{2h-1}{(z-w)^3}
+\frac{N(w)}{(z-w)^2}
+\frac{\partial N(w)}{z-w}.
\]
Since \(h=2\) for \((\widetilde b,\widetilde c)\) and \(S^{\gh}\)
contains \(2N_M\), the mixed anomaly is
\[
c_{LS}^{\gh}=2(2h-1)\big|_{h=2}=2\times3=6.
\]
Finally, because
\[
N_M(z)N_M(w)\sim\frac{1}{(z-w)^2},
\]
the normalization \(S^{\gh}\supset2N_M\) gives
\[
c_{SS}^{\gh}=2^2\times1=4.
\]
Hence,
\[
\left(c_{LL},c_{LS},c_{SS}\right)_{\gh}
=
(-54,6,4).
\]

Two checks protect this result from normalization ambiguities.  First,
in the unit-charge basis \(\mathcal D=S/2\), the same ghost complex has
\((c_{L\mathcal D},c_{\mathcal D\mathcal D})=(3,1)\)
\cite{Batlle2024}; rescaling back multiplies these coefficients by two
and four.  Second, the current \(M^\gh\) contains
\(\widetilde b\) but no \(\widetilde c\).  It follows before any
regularization that \(M^\gh M^\gh\) has no central contraction, in
agreement with the abstract algebra.\footnote{In the physical constraint basis the ghost currents are, by the same
dictionary as in the matter sector,
\begin{equation*}
 \mathcal C_1^\gh=2M^\gh,\qquad
 \mathcal C_2^\gh=-T^\gh,\qquad
 \mathcal C_3^\gh=S^\gh .
\label{eq:physical-ghost-currents}
\end{equation*}
It follows immediately that
\begin{equation*}
 (k_{22},k_{23},k_{33})_\gh=(-54,-6,4).
\label{eq:physical-ghost-anomalies}
\end{equation*}
The sign of \(k_{23}\) is the sign from
\(\mathcal C_2=-T\); its zero and hence the nilpotency condition are
basis independent.}

\subsection*{Total current algebra}

Matter and ghosts are independent, so their central terms add.  Define
\begin{equation}
 T^\tot=T^X+T^\gh,\qquad
 M^\tot=M^X+M^\gh,\qquad
 S^\tot=S^X+S^\gh.
\label{eq:total-currents}
\end{equation}
At equal Carrollian time,
\begin{align}
 T^\tot(z)T^\tot(w)
 &\sim
 \frac{2D-54}{2(z-w)^4}
 +\frac{2T^\tot(w)}{(z-w)^2}
 +\frac{\partial T^\tot(w)}{z-w},
\label{eq:TtotTtot}\\
 T^\tot(z)S^\tot(w)
 &\sim
 \frac{6-D}{(z-w)^3}
 +\frac{S^\tot(w)}{(z-w)^2}
 +\frac{\partial S^\tot(w)}{z-w},
\label{eq:TtotStot}\\
 S^\tot(z)S^\tot(w)
 &\sim\frac{4-D}{(z-w)^2},
\label{eq:StotStot}\\
 T^\tot(z)M^\tot(w)
 &\sim
 \frac{2M^\tot(w)}{(z-w)^2}
 +\frac{\partial M^\tot(w)}{z-w},
\label{eq:TtotMtot}\\
 S^\tot(z)M^\tot(w)
 &\sim\frac{2M^\tot(w)}{z-w},
\qquad
 M^\tot(z)M^\tot(w)\sim0.
\label{eq:remaining-total-opes}
\end{align}
Equivalently, define the total physical constraint currents
\begin{equation}
 \mathcal C_i^\tot=\mathcal C_i^X+\mathcal C_i^\gh,\qquad
 \mathcal C_1^\tot=2M^\tot,\quad
 \mathcal C_2^\tot=-T^\tot,\quad
 \mathcal C_3^\tot=S^\tot .
\label{eq:physical-total-currents}
\end{equation}
Their anomalous products are
\begin{align}
 \mathcal C_2^\tot(z)\mathcal C_2^\tot(w)
 &\sim\frac{2D-54}{2u^4}
 -\frac{2\mathcal C_2^\tot(w)}{u^2}
 -\frac{\partial\mathcal C_2^\tot(w)}{u},
\label{eq:C2C2-total}\\
 \mathcal C_2^\tot(z)\mathcal C_3^\tot(w)
 &\sim\frac{D-6}{u^3}
 -\frac{\mathcal C_3^\tot(w)}{u^2}
 -\frac{\partial\mathcal C_3^\tot(w)}{u},
\label{eq:C2C3-total}\\
 \mathcal C_3^\tot(z)\mathcal C_3^\tot(w)
 &\sim\frac{4-D}{u^2}.
\label{eq:C3C3-total}
\end{align}
Thus the anomaly vector in the convention of
eq.~\eqref{eq:first-principles-C} is
\begin{equation}
 (k_{22},k_{23},k_{33})_\tot
 =(2D-54,D-6,4-D).
\label{eq:physical-total-anomaly-vector}
\end{equation}
The anomaly bookkeeping is summarized in \cref{tab:anomalies}.

\begin{table}[H]
\centering
\renewcommand{\arraystretch}{1.22}
\begin{tabular}{|>{\raggedright\arraybackslash}p{0.25\textwidth}|
                 >{\centering\arraybackslash}p{0.18\textwidth}|
                 >{\centering\arraybackslash}p{0.18\textwidth}|
                 >{\centering\arraybackslash}p{0.18\textwidth}|}
\hline
Sector & \(\cL\) & \(\cS\) & \(\cSS\) \\
\hline
Matter \(X,P\) & \(2D\) & \(-D\) & \(-D\) \\
Ghosts \(b,c,\widetilde b,\widetilde c,r,s\)
 & \(-54\) & \(6\) & \(4\) \\
\hline
Total & \(2D-54\) & \(6-D\) & \(4-D\) \\
\hline
\end{tabular}
\caption{The three cohomologically independent anomaly coefficients in
the standard \(L=-C_2\), \(M=C_1/2\), \(S=C_3\) basis. In the
physical constraint basis the mixed column changes sign, as displayed
in eq.~\eqref{eq:physical-total-anomaly-vector}.}
\label{tab:anomalies}
\end{table}

\section{BRST charge, nilpotency, and the critical dimension test}
\label{sec:brst}

Having determined the total current algebra, we can now formulate the
consistency test invariantly.  The question is not whether one preferred
central charge vanishes, but whether every cocycle that can appear in
the square of the BRST charge cancels simultaneously.

\subsection*{Classical BFV differential}

Let \(c,\widetilde c,s\) denote the ghosts for the standard-basis
generators \(L,M,S\), respectively.  The ghosts for the exact
constraints \(C_i\) are the \(\gamma_i\) of
eq.~\eqref{eq:constraint-ghost-dictionary}.
The centerless algebra determines their differential through 
\begin{equation}
 \delta_B c^A=-\frac12f_{BC}{}^A c^Bc^C,
\label{eq:CE-differential}
\end{equation}
including the mode-dependent structure constants generated by spatial
derivatives.  Here a condensed index includes both the constraint
species and its Fourier mode.  The corresponding classical BFV charge
is
\begin{equation}
 Q_{\mathrm{BFV}}=Q_1+Q_2,\qquad
 Q_1=c^AG_A,\qquad
 Q_2=-\half f_{AB}{}^Cc^Ac^Bb_C .
\label{eq:classical-BFV-split}
\end{equation}
The square of the linear term is
\begin{equation}
 Q_1^2
 =\half c^Ac^B[G_A,G_B]
 =\half c^Ac^Bf_{AB}{}^CG_C .
\label{eq:classical-Q1-square}
\end{equation}
This constraint-valued expression cancels against the corresponding
part of \(\{Q_1,Q_2\}\).  The uncancelled three-ghost terms are
proportional to
\begin{equation}
 c^Ac^Bc^C f_{[AB}{}^D f_{C]D}{}^E b_E,
\label{eq:classical-BFV-Jacobi}
\end{equation}
which vanishes by the Jacobi identity.  Thus the centerless classical
charge is nilpotent before any equations of motion or physical-state
conditions are imposed.

In local notation the same differential gives
eqs.~\eqref{eq:brst-c}, \eqref{eq:brst-s}, and \eqref{eq:brst-ctilde}.  Their individual terms have
a direct origin
\begin{align}
 c\partial c
 &\longleftrightarrow [L,L]\sim L,
\nonumber\\
 c\partial s
 &\longleftrightarrow [L,S]\sim S,
\nonumber\\
 c\partial\widetilde c+\widetilde c\partial c
 &\longleftrightarrow [L,M]\sim M,
\qquad
 2s\widetilde c
 \longleftrightarrow [S,M]=2M.
\label{eq:ghost-structure-map}
\end{align}
Using the graded Leibniz rule, one verifies
\begin{equation}
 \delta_B^2c=\delta_B^2s=\delta_B^2\widetilde c=0
\label{eq:classical-ghost-nilpotency}
\end{equation}
without equations of motion.  For \(\widetilde c\), the terms involving
\(s\widetilde c\partial c\) cancel only with the physical coefficient
two.  This is a local ghost-level check of the Jacobi identities.

\subsection{The local quantum current}

The minimal BRST current can be written directly from
eqs.~\eqref{eq:brst-c}, \eqref{eq:brst-s}, and \eqref{eq:brst-ctilde}:
\begin{align}
 j_B={}&
 \normal{cT^X}+\normal{sS^X}+\normal{\widetilde c M^X}
 +\normal{bc\partial c}
 +\normal{r c\partial s}
\nonumber\\
 &+\normal{\widetilde b c\partial\widetilde c}
 +\normal{\widetilde b\widetilde c\partial c}
 +2\normal{s\widetilde b\widetilde c}.
\label{eq:BRST-current-explicit}
\end{align}
Up to a total derivative it takes the symmetric form
\begin{equation}
 j_B=
 \normal{cT^X}+\normal{sS^X}+\normal{\widetilde c M^X}
 +\half\left(
 \normal{cT^\gh}+\normal{sS^\gh}
 +\normal{\widetilde c M^\gh}
 \right).
\label{eq:BRST-current-symmetric}
\end{equation}
Using the exact-constraint ghosts in
eq.~\eqref{eq:constraint-ghost-dictionary}, the same current is
\begin{equation}
 j_B=
 \sum_{i=1}^3\normal{\gamma_i\mathcal C_i^X}
 +\frac12\sum_{i=1}^3\normal{\gamma_i\mathcal C_i^\gh}
 +\partial(\text{ghost-number-one field}).
\label{eq:BRST-current-physical-basis}
\end{equation}
Equation~\eqref{eq:BRST-current-physical-basis} makes explicit that the
charge couples to \(C_1=P^2\), \(C_2=P\cdot X'\), and
\(C_3=P\cdot X\), not to a different constraint surface.  The final
term denotes the same harmless total-derivative freedom already
present in eq.~\eqref{eq:BRST-current-symmetric}.
The BRST charge is
\begin{equation}
 Q_B=\oint\frac{\dd z}{2\pi\ii}\,j_B(z).
\label{eq:BRST-charge}
\end{equation}
Its OPEs with the antighosts return the total constraints
\begin{equation}
 \{Q_B,b\}=T^\tot,\qquad
 \{Q_B,\widetilde b\}=M^\tot,\qquad
 \{Q_B,r\}=S^\tot.
\label{eq:Q-antighost}
\end{equation}
Equivalently, in the physical constraint basis,
\begin{equation}
 \{Q_B,\beta^i\}=\mathcal C_i^\tot,\qquad i=1,2,3.
\label{eq:Q-antighost-physical}
\end{equation}
This is the precise counterpart of
\(\{Q_B,b\}=T^{\mat}+T^\gh\) in the ordinary bosonic string.

The factor \(1/2\) in eq.~\eqref{eq:BRST-current-symmetric} prevents double
counting the adjoint ghost contribution.  Expanding the right-hand side
and integrating total derivatives around the contour reproduces
eq.~\eqref{eq:BRST-current-explicit}.  The current has conformal weight one and
ghost number one.  Possible quantum improvements are derivatives of
ghost-number-one, weight-zero operators.  Such improvements can change
the local representative of \(j_B\), but their contour integrals vanish
and they cannot alter a nontrivial central cocycle.

Applying \(Q_B\) once more to eq.~\eqref{eq:Q-antighost} gives
\begin{equation}
 \{Q_B,\{Q_B,b_A\}\}
 =\{Q_B,G_A^{\tot}\}.
\label{eq:double-on-antighost}
\end{equation}
If the total algebra is centerless, the right-hand side is the
constraint-algebra action on \(G_A^{\tot}\) and cancels by the same
Jacobi identities as at the classical level.  A central term has no
constraint on which the remaining ghost can act, so it survives.  This
is why central-charge cancellation is a necessary condition for quantum
nilpotency.

\subsection{The allowed central extensions}

Before imposing nilpotency, it is useful to know which anomaly channels
are intrinsic to the algebra.  The most general local central extension
compatible with the current weights can be put in the form
\begin{align}
 [L_m,L_n]&=(m-n)L_{m+n}
 +\frac{\cL}{12}m(m^2-1)\delta_{m+n,0},
\label{eq:mode-LL}\\
 [L_m,M_n]&=(m-n)M_{m+n},
\label{eq:mode-LM}\\
 [L_m,S_n]&=-nS_{m+n}
 +\frac{\cS}{2}m(m+1)\delta_{m+n,0},
\label{eq:mode-LS}\\
 [S_m,M_n]&=2M_{m+n},
\label{eq:mode-SM}\\
 [S_m,S_n]&=\cSS\,m\delta_{m+n,0},
\qquad [M_m,M_n]=0.
\label{eq:mode-SS}
\end{align}
These are the Virasoro, mixed Virasoro-current, and affine-current
cocycles of the Weyl-BMS algebra \cite{Batlle2024}.

\subsection*{Absence of an independent \(LM\) central extension}

The ordinary two-generator BMS algebra may admit a central extension in
the mixed \(LM\) commutator.  In the Carroll-Weyl algebra, however,
the additional generator \(S_m\) forbids such an extension.  The
relevant centerless commutators are
\begin{align}
 [L_m,M_n] &= (m-n)M_{m+n}, \label{eq:LM-centerless}\\
 [L_m,S_n] &= -nS_{m+n}, \label{eq:LS-centerless}\\
 [S_m,M_n] &= 2M_{m+n}. \label{eq:SM-centerless}
\end{align}
Setting the mode of the Carroll-Weyl generator to zero gives
\begin{equation}
 [S_0,L_m]=0,
 \qquad
 [S_0,M_n]=2M_n.
 \label{eq:S0-actions}
\end{equation}
Thus \(L_m\) is neutral under \(S_0\), whereas \(M_n\) has
Carroll-Weyl charge two.

Suppose, provisionally, that the \(LM\) commutator contains an
independent central term
\begin{equation}
 [L_m,M_n]
 =
 (m-n)M_{m+n}
 +
 K_{LM}(m)\delta_{m+n,0},
 \label{eq:LM-hypothetical}
\end{equation}
where \(K_{LM}(m)\) is a central polynomial. By definition, the
central element commutes with every generator and, in particular, with
\(S_0\).

We now impose the Jacobi identity for the triple
\((S_0,L_m,M_n)\),
\begin{equation}
 [S_0,[L_m,M_n]]
 +
 [L_m,[M_n,S_0]]
 +
 [M_n,[S_0,L_m]]
 =0.
 \label{eq:S0LM-Jacobi}
\end{equation}

The three terms can be evaluated separately.  Using
eqs.~\eqref{eq:LM-hypothetical} and \eqref{eq:S0-actions}, the first
term is
\begin{align}
 [S_0,[L_m,M_n]]
 &=
 [S_0,(m-n)M_{m+n}
       +K_{LM}(m)\delta_{m+n,0}]
 \nonumber\\
 &=
 2(m-n)M_{m+n}.
 \label{eq:first-Jacobi-term}
\end{align}
There is no contribution from the central term because
\begin{equation}
 [S_0,K_{LM}(m)\delta_{m+n,0}]=0.
\end{equation}

For the second term, antisymmetry gives
\begin{equation}
 [M_n,S_0]=-2M_n,
\end{equation}
and therefore
\begin{align}
 [L_m,[M_n,S_0]]
 &=
 -2[L_m,M_n]
 \nonumber\\
 &=
 -2(m-n)M_{m+n}
 -2K_{LM}(m)\delta_{m+n,0}.
 \label{eq:second-Jacobi-term}
\end{align}

The third term vanishes,
\begin{equation}
 [M_n,[S_0,L_m]]=0,
 \label{eq:third-Jacobi-term}
\end{equation}
because \([S_0,L_m]=0\).

Substituting
eqs.~\eqref{eq:first-Jacobi-term}--\eqref{eq:third-Jacobi-term}
into the Jacobi identity \eqref{eq:S0LM-Jacobi}, the non-central
terms cancel
\begin{align}
 0
 &=
 2(m-n)M_{m+n}
 -2(m-n)M_{m+n}
 -2K_{LM}(m)\delta_{m+n,0}
 \nonumber\\
 &=
 -2K_{LM}(m)\delta_{m+n,0}.
 \label{eq:LM-Jacobi}
\end{align}
Hence
\begin{equation}
K_{LM}(m)=0.
 \label{eq:LM-cocycle-vanishes}
\end{equation}

The result can also be understood directly from Carroll-Weyl charge.
The operator \(M_n\) has charge two under \(S_0\), so the entire
right-hand side of \([L_m,M_n]\) must transform with the same charge.
The non-central term \(M_{m+n}\) has precisely this charge.  A central
element, however, commutes with \(S_0\) and therefore has charge zero.
A sum of a charge-two operator and a charge-zero central element cannot
transform homogeneously with charge two.  The Jacobi identity enforces
this charge mismatch by setting the central coefficient to zero.

Equivalently, in the local current algebra, a hypothetical \(LM\)
central extension would appear as a field-independent fourth-order
pole,
\begin{equation}
 T(z)M(w)
 \supset
 \frac{c_{LM}}{(z-w)^4}.
 \label{eq:TM-hypothetical-pole}
\end{equation}
However, the Carroll-Weyl current acts on \(M\) according to
\begin{equation}
 S(z)M(w)\sim \frac{2M(w)}{z-w},
\end{equation}
while a field-independent pole carries zero Carroll-Weyl charge.
The corresponding Ward identity is therefore inconsistent unless
\begin{equation}
 c_{LM}=0.
\end{equation}

Thus the usual mixed BMS cocycle cannot be extended to the complete
Carroll-Weyl algebra.  The \(LM\) commutator retains its non-central
part,
\begin{equation}
 [L_m,M_n]=(m-n)M_{m+n},
\end{equation}
but it possesses no independent central charge.  The only independent
central extensions relevant to BRST nilpotency are consequently the
\(LL\), \(LS\), and \(SS\) cocycles. The Carroll-Weyl extension removes the usual
\(c_M\) cocycle of the BMS subalgebra \cite{Batlle2024}.

\subsection{Nilpotency conditions}

\subsubsection*{Operator meaning of nilpotency}

The statement \(Q_B^2=0\) is an operator identity on the entire BRST
complex; it is not a condition imposed only on a preferred set of
states.  Since \(Q_B\) is Grassmann odd and has ghost number one, its
square and its action on a homogeneous operator \(\mathcal O\) are
\begin{equation}
 Q_B^2=\half\{Q_B,Q_B\},\qquad
 \delta_B\mathcal O=[Q_B,\mathcal O\}_{\mathrm{gr}},\qquad
 \delta_B^2\mathcal O=[Q_B^2,\mathcal O\}_{\mathrm{gr}} .
\label{eq:BRST-square-meaning}
\end{equation}
Thus a nilpotent charge defines a differential.  More importantly, the
physical space is meant to be the cohomology
\begin{equation}
 \cH_{\mathrm{phys}}
 =\frac{\ker Q_B}{\operatorname{im}Q_B}.
\label{eq:BRST-cohomology-definition}
\end{equation}
This quotient exists only when
\(\operatorname{im}Q_B\subseteq\ker Q_B\).  Indeed, for
\(Q_B|\Psi\rangle=0\), a putative gauge-equivalent representative obeys
\begin{equation}
 Q_B\bigl(|\Psi\rangle+Q_B|\Lambda\rangle\bigr)
 =Q_B^2|\Lambda\rangle .
\label{eq:BRST-exact-shift-test}
\end{equation}
For an arbitrary gauge parameter \(|\Lambda\rangle\), the right-hand
side vanishes precisely when \(Q_B^2\) vanishes as an operator.
Consequently, a nonzero \(Q_B^2\) destroys the equivalence relation
used to define physical states; it is not merely a shift of a
mass-shell or intercept condition.

\subsubsection*{From a central extension to the BRST obstruction}

The mechanism is transparent in the general BFV construction.  If
quantum normal ordering changes the constraint algebra to
\begin{equation}
 [G_A,G_B]
 =f_{AB}{}^C G_C+K_{AB}\,\mathbf 1 ,
\label{eq:quantum-BFV-algebra}
\end{equation}
the part of the BRST charge linear in the constraints gives
\begin{equation}
 \half c^Ac^B[G_A,G_B]
 =\half c^Ac^Bf_{AB}{}^CG_C
 +\half c^Ac^BK_{AB}.
\label{eq:BFV-central-split}
\end{equation}
The first term is cancelled by the linear-cubic and cubic-cubic
ghost contributions: this is the usual structure-constant cancellation,
and the residue is proportional to the Jacobi identity.  The second
term contains no constraint generator and survives as a ghost-number-two
operator.  In the formulas below the normal-ordering contributions of
all three ghost systems have already been included in
\(\cL^\tot,\cS^\tot,\cSS^\tot\); there is therefore no further hidden
ghost term available to cancel these coefficients.

Equivalently, applying \(Q_B\) twice to each identity in
eq.~\eqref{eq:Q-antighost} shows that nilpotency requires the quantum
total constraints to represent the centerless algebra.  Evaluating the
central part explicitly gives
\begin{align}
 Q_B^2\big|_{\mathrm{central}}
 ={}&
 \frac{\cL^\tot}{24}
 \sum_m m(m^2-1)c_{-m}c_m
 +\frac{\cS^\tot}{4}
 \sum_m m(m+1)
 \left(c_{-m}s_m-s_{-m}c_m\right)
\nonumber\\
 &+\frac{\cSS^\tot}{2}
 \sum_m m\,s_{-m}s_m
 +Q_{\mathrm{zero\ mode}}^2 .
\label{eq:Q2-central}
\end{align}
The last term denotes the separate zero-mode or intercept contribution.
It cannot contain any of the nonzero-mode monomials displayed in the
first three terms.

\subsubsection*{Linear independence of the three anomaly terms}

Pairing the \(m=p\) and \(m=-p\) terms in
eq.~\eqref{eq:Q2-central}, and using the fermionic anticommutation of
the \(c\) and \(s\) oscillators, gives the equivalent positive-mode
form
\begin{align}
 Q_B^2\big|_{\mathrm{central}}
 ={}&
 \frac{\cL^\tot}{12}
 \sum_{p=1}^{\infty}p(p^2-1)c_{-p}c_p
\nonumber\\
 &+\frac{\cS^\tot}{2}
 \sum_{p=1}^{\infty}p^2
 \bigl(c_{-p}s_p-s_{-p}c_p\bigr)
\nonumber\\
 &+\cSS^\tot\sum_{p=1}^{\infty}p\,s_{-p}s_p
 +Q_{\mathrm{zero\ mode}}^2 .
\label{eq:Q2-central-positive}
\end{align}
Although the three displayed terms all have total ghost number two,
they have different \((c,s)\)-species content,
\begin{equation}
 c_{-p}c_p:(2,0),\qquad
 c_{-p}s_p-s_{-p}c_p:(1,1),\qquad
 s_{-p}s_p:(0,2).
\label{eq:ghost-bidegrees}
\end{equation}
The free ghost algebra is an exterior algebra in the independent
\(c_m\) and \(s_m\) generators.  Monomials with different species
bidegree, or with different mode labels, form independent basis
elements.  This gives an algebraic proof that no numerical cancellation
between the three sums is possible.

The conclusion can be seen already in the \(p=2\) component,
\begin{equation}
 Q_B^2\big|_{\mathrm{central},\,p=2}
 =\frac{\cL^\tot}{2}c_{-2}c_2
 +2\cS^\tot\bigl(c_{-2}s_2-s_{-2}c_2\bigr)
 +2\cSS^\tot s_{-2}s_2 .
\label{eq:Q2-p-two}
\end{equation}
Two \(b\)-antighost probes isolate the \(cc\) monomial, one \(b\) and
one \(r\) isolate the \(cs\) monomial, and two \(r\)'s isolate the
\(ss\) monomial, because
\(\{b_m,c_n\}=\{r_m,s_n\}=\delta_{m+n,0}\) while all cross-species
anticommutators vanish.  Hence the operator identity \(Q_B^2=0\)
forces each coefficient in eq.~\eqref{eq:Q2-p-two} to vanish,
\begin{equation}
 Q_B^2=0
 \quad\Longrightarrow\quad
 \cL^\tot=0,\qquad
 \cS^\tot=0,\qquad
 \cSS^\tot=0,
\label{eq:nilpotency-three}
\end{equation}
together with a separate zero-mode or intercept condition.
Conversely, once the three nontrivial cocycles and the zero-mode
obstruction vanish, the remaining terms cancel by the classical BFV
and Jacobi identities.  Thus, within the present minimal complex,
\begin{equation}
 Q_B^2=0
 \quad\Longleftrightarrow\quad
 \cL^\tot=\cS^\tot=\cSS^\tot=0
 \quad\text{and}\quad
 Q_{\mathrm{zero\ mode}}^2=0 .
\label{eq:nilpotency-equivalence}
\end{equation}
In particular, a relation such as
\(\cL^\tot+\cS^\tot+\cSS^\tot=0\) would be irrelevant: it would only
cancel three numbers, whereas the numbers multiply three different
operators.

In the exact constraint basis,
\(c=-\gamma_2\) and \(s=\gamma_3\).  The three independent monomials
are therefore \(\gamma_2\gamma_2\),
\(\gamma_2\gamma_3\), and \(\gamma_3\gamma_3\), multiplied by
\(k_{22}\), \(k_{23}\), and \(k_{33}\), respectively.  Using
eq.~\eqref{eq:physical-total-anomaly-vector}, the three equations are
\begin{equation}
 2D-54=0,\qquad D-6=0,\qquad 4-D=0.
\label{eq:three-D-equations}
\end{equation}
The standard-basis form of the middle equation is \(6-D=0\); the two
expressions differ only because \(L=-C_2\).
They separately demand
\begin{equation}
 D_{C_2C_2}=27,\qquad D_{C_2C_3}=6,\qquad D_{C_3C_3}=4.
\label{eq:three-D-values}
\end{equation}
There is no common target-space dimension.  This is the central result
of the paper. As shown in Table~\ref{tab:special-D}, each of the dimensions \(D=27,6,4\) cancels exactly one of the three anomaly coefficients.

\begin{table}[H]
\centering
\renewcommand{\arraystretch}{1.17}
\begin{tabular}{|c|c|c|c|}
\hline
 \(D\) & \(\cL^\tot\) & \(\cS^\tot\) & \(\cSS^\tot\) \\
\hline
\(27\) & \(0\)   & \(-21\) & \(-23\) \\
\(6\)  & \(-42\) & \(0\)   & \(-2\)  \\
\(4\)  & \(-46\) & \(2\)   & \(0\)   \\
\hline
\end{tabular}
\caption{Vanishing one cocycle leaves the other two anomalous. The
numbers \(27,6,4\) are roots of separate consistency conditions, not
three alternative critical dimensions.}
\label{tab:special-D}
\end{table}

\subsection*{Direct contour interpretation of the three terms}

The same implication follows locally, without choosing a ghost Fock
vacuum.  Since \(Q_B\) is odd, its square is the nested contour
\begin{equation}
 Q_B^2
 =\half\oint\frac{\dd w}{2\pi\ii}
 \oint_{z=w}\frac{\dd z}{2\pi\ii}\,
 j_B(z)j_B(w).
\label{eq:BRST-double-contour}
\end{equation}
After the non-central BFV cancellations, only the three anomalous poles
of the total currents can contribute,
\begin{align}
 T^\tot(z)T^\tot(w)\big|_{\mathrm{cen}}
 &\sim\frac{\cL^\tot/2}{(z-w)^4},
\nonumber\\
 T^\tot(z)S^\tot(w)\big|_{\mathrm{cen}}
 &\sim\frac{\cS^\tot}{(z-w)^3},
\nonumber\\
 S^\tot(z)S^\tot(w)\big|_{\mathrm{cen}}
 &\sim\frac{\cSS^\tot}{(z-w)^2}.
\label{eq:central-poles-for-Q2}
\end{align}
The fourth-order pole multiplies \(c(z)c(w)\).  In the Taylor expansion
\begin{equation}
 c(z)=c(w)+(z-w)\partial c(w)
 +\frac{(z-w)^2}{2}\partial^2c(w)
 +\frac{(z-w)^3}{6}\partial^3c(w)+\cdots,
\label{eq:c-Taylor-for-Q2}
\end{equation}
the cubic term is the one that produces the residue, leaving
\(c\partial^3c\).  The two radial orderings of the mixed products
\(cT\) and \(sS\) probe the third-order \(TS\) pole.  Taylor expanding
both ghosts and combining the two orderings leaves, up to a total
derivative, the independent local operator
\begin{equation}
 c\,\partial^2s-s\,\partial^2c
\label{eq:local-cs-anomaly}
\end{equation}
Finally, the term linear in \((z-w)\) in \(s(z)s(w)\) combines with the
second-order \(SS\) pole to leave \(s\partial s\).  Consequently the
local obstruction has the schematic form
\begin{align}
 Q_B^2\big|_{\mathrm{central}}
 \sim{}&
 \cL^\tot\oint c\,\partial^3c
\nonumber\\
 &+\cS^\tot\oint
 \bigl(c\,\partial^2s-s\,\partial^2c\bigr)
\nonumber\\
 &+\cSS^\tot\oint s\,\partial s .
\label{eq:local-Q2-obstruction}
\end{align}
The nonzero numerical factors suppressed in eq.~\eqref{eq:local-Q2-obstruction}
are displayed exactly in eq.~\eqref{eq:Q2-central}.  The three local
operators again have species bidegrees \((2,0)\), \((1,1)\), and
\((0,2)\), so integrations by parts or local improvements cannot turn
one into another.  This contour argument is the local counterpart of
the oscillator projection in eq.~\eqref{eq:Q2-p-two}.

The \(M\)-ghost does not appear in the central part of \(Q_B^2\).
This is not because the \(M\) constraint is dispensable: its ghost and
antighost occur essentially in \(S^\gh\), \(M^\gh\), and the cubic
BRST vertex.  Rather, the \(TM\), \(SM\), and \(MM\) OPEs carry no
independent cocycles in the gauge-complete algebra.

\subsection*{Robustness under intercept shifts and current rescalings}

An intercept changes \(L_0\) by a constant,
\(L_0\mapsto L_0-a\).  It can modify a term linear in \(m\) in the
\(LL\) commutator, but it cannot alter the coefficient of \(m^3\).
Likewise, shifting \(S_0\) can change a coboundary term linear in \(m\)
in the \(LS\) bracket, but not its quadratic cocycle.  Neither operation
changes the \(SS\) affine level.  The three nontrivial coefficients in
eq.~\eqref{eq:three-D-equations} must therefore vanish independently.

One might instead rescale the physical current, \(S\to\alpha S\).
The matter and ghost mixed coefficients both scale as \(\alpha\), and
the affine levels both scale as \(\alpha^2\).  Thus the conditions
\(D=6\) and \(D=4\) are invariant under a nonzero change of
normalization.  Finally, an improvement \(T\to T+\beta\partial S\)
changes the conformal weight of \(M\), because
\(\partial S(z)M(w)\) contains a second-order pole.  For the fixed
classical algebra in eq.~\eqref{eq:centerless-LMS}, \(\beta\) must vanish.

\subsection*{Conditional physical-state conditions}

Although the anomaly prevents a nilpotent quantum cohomology in the
minimal highest-weight theory, it is useful to record what the relative
BRST conditions would be if the three cocycles were cancelled by an
enlarged matter sector.  Choosing the relative complex,
\begin{equation}
 b_0|\Psi\rangle=\widetilde b_0|\Psi\rangle
 =r_0|\Psi\rangle=0,
\label{eq:relative-complex}
\end{equation}
the antighost identities imply
\begin{equation}
 L_n^{\tot}|\Psi\rangle
 =M_n^{\tot}|\Psi\rangle
 =S_n^{\tot}|\Psi\rangle=0
 \qquad(n>0),
\label{eq:positive-mode-physical}
\end{equation}
together with zero-mode mass-shell, level, and scaling conditions.  The
\(S_0\) equation is new relative to the traditional null string: it
requires states or vertex operators to carry a definite compensating
homogeneity under target-space rescalings generated by \(X\cdot P\).

One must not impose eq.~\eqref{eq:positive-mode-physical} and then call the
result a quantum BRST spectrum when \(Q_B^2\neq0\).  If
\(|\Psi\rangle\sim|\Psi\rangle+Q_B|\Lambda\rangle\), nilpotency is what
guarantees that a BRST-exact shift preserves \(Q_B|\Psi\rangle=0\).
The anomaly therefore obstructs the equivalence relation defining the
cohomology, not merely a particular mass-shell condition.

\section{Interpretation and relation to other quantizations}
\label{sec:interpret}

The three roots \(27,6,4\) acquire meaning only after the gauge complex
and representation are specified.  We therefore compare complete
quantization problems, rather than transferring a ``critical
dimension'' from one null-string formulation to another.

\subsection*{The phrase ``critical dimension'' labels different tests}

Several quantum theories associated with a tensionless or null
worldsheet are close enough to invite comparison but differ in their
constraints, ghosts, and representation. The earliest covariant
quantizations of the null string already disagreed on whether a critical
dimension was present \cite{Lizzi1986,Bozhilov1997}. The conformal string adds target-space conformal constraints; its Hamiltonian BRST analysis yields differing results in the literature, selecting \(D=2\) in \cite{Gustafsson1995}, but \(D=6\) and \(D=26\) in \cite{Chen:2026cau}. Recently, it was shown that the Carroll-Weyl gauged null string in
$d$ dimensions can be obtained by a Dirac reduction of the conformal null
string formulated in Dirac's $(d+2)$-dimensional embedding space \cite{Lindstrom:2026quz, Lindstrom:2026zno}. In
particular, the reduction maps the semidirect product of the Virasoro
algebra with the $\mathfrak{su}(1,1)$ Kac-Moody algebra to the
Carroll-Weyl constraint algebra of the $d$-dimensional theory
\cite{Lindstrom:2026zno}. Ambitwistor
strings are chiral theories of maps to null geodesics with their own
gauge complex; the bosonic model has the standard \(D=26\) critical
dimension
\cite{MasonSkinner2014,CasaliTourkine2016,CasaliHerfrayTourkine2017}.

Table~\ref{tab:critical-comparison} compares the dimension tests in four
quantizations and shows how the inclusion of \(S\) changes the result
for the highest-weight null string.

\begin{table}[H]
\centering
\small
\setlength{\tabcolsep}{4pt}
\renewcommand{\arraystretch}{1}
\begin{tabular}{|>{\raggedright\arraybackslash}p{0.23\textwidth}|
                 >{\raggedright\arraybackslash}p{0.29\textwidth}|
                 >{\raggedright\arraybackslash}p{0.18\textwidth}|
                 >{\raggedright\arraybackslash}p{0.18\textwidth}|}
\hline
Quantization & Defining quantum data & Reported dimension test &
Relation to this work \\
\hline
Tensile bosonic string
& Two Virasoro algebras and two \(bc\) systems
& \(D=26\) & Different, non-degenerate worldsheet \\
Traditional highest-weight null string
& \(L,M\) constraints and two BMS \(bc\)-ghost pairs
& \(D=26\) & Partially gauge-fixed subsector \\
Bosonic ambitwistor string
& Chiral null-geodesic constraints
& \(D=26\) & Related null geometry, different gauge system \\
\hline
Carroll--Weyl highest-weight string
& \(\begin{array}{@{}c@{}}
      (L,M) \\[-2pt]
      \downarrow \\[-2pt]
      (L,M,S)
    \end{array}\)
  \newline Three mixed ghost pairs
& No common \(D\) & Present calculation \\
\hline
\end{tabular}
\caption{Critical-dimension statements are properties of a complete
quantum gauge complex and its representation, not of the word
``tensionless'' by itself.}
\label{tab:critical-comparison}
\end{table}

\subsection*{Enlargement of the old \texorpdfstring{\(D=26\)}{D=26} check}

If the Carroll-Weyl constraint and the \((r,s)\) pair are deleted, the
ghost sector contains two weight-\((2,-1)\) pairs.  Then
\begin{equation}
 \cL^{\gh,\BMS}=-26-26=-52,
\end{equation}
while \(D\) first-order matter pairs give \(\cL^X=2D\).  The single
remaining highest-weight anomaly condition is
\begin{equation}
 2D-52=0\qquad\Longrightarrow\qquad D=26,
\label{eq:old-D26}
\end{equation}
which reproduces the intrinsic BMS ghost calculation of
\cite{Chen2023}.  In the gauge-complete theory, adding only the scalar
ghost central charge changes this \(TT\) condition to \(D=27\), but that
number has no independent claim to criticality: at \(D=27\),
\begin{equation}
 \cS^\tot=-21,\qquad \cSS^\tot=-23.
\end{equation}
The new constraint turns a one-anomaly problem into a three-anomaly
problem.

It is illuminating to perform the operations in two stages.  Starting
with the old BMS complex at \(D=26\), adding the scalar \((r,s)\) pair
changes the Virasoro coefficient by \(-2\), so the \(TT\) channel is no
longer cancelled.  But the semidirect completion simultaneously creates
the \(LS\) and \(SS\) ghost anomalies \(6\) and \(4\), while the matter
sector contributes \(-26\) to each.  The failure is therefore not a
small shift from \(26\) to \(27\); it is the appearance of two new,
large uncancelled cocycles.

\subsection*{Agreement with the abstract Weyl-BMS BRST complex}

The numerical result has a useful algebraic cross-check.  Let
\(\mathcal D=S/2\), so that
\([\mathcal D_m,M_n]=M_{m+n}\).  The \(\lambda=-1\) Weyl-BMS ghost
complex of \cite{FigueroaVishwa2025,Batlle2024} has
\begin{equation}
 (c_L,c_{L\mathcal D},c_{\mathcal D\mathcal D})_\gh
 =(-54,3,1).
\label{eq:unit-charge-ghost}
\end{equation}
The first-order matter realization has
\begin{equation}
 (c_L,c_{L\mathcal D},c_{\mathcal D\mathcal D})_X
 =\left(2D,-\frac{D}{2},-\frac{D}{4}\right).
\label{eq:unit-charge-matter}
\end{equation}
Restoring \(S=2\mathcal D\) multiplies the mixed coefficient by two
and the affine level by four, giving exactly
eqs.~\eqref{eq:matter-anomaly-vector} and \eqref{eq:ghost-anomaly-vector}.  The present
calculation therefore identifies the gauge-fixed Carroll-Weyl null
string as a concrete sigma-model realization of a Weyl-BMS BRST
complex whose matter charges do not meet the critical values.

\subsection*{Vacuum dependence and the light-cone result}
\label{subsec:vacuum}

An OPE is not determined by the equal-time canonical brackets alone.
It also requires a splitting into creation and annihilation parts.  Our
use of eqs.~\eqref{eq:matter-basic-ope} and \eqref{eq:ghost-basic-ope} selects a
highest-weight, or flipped, representation.  This is the representation
in which the normal-ordered current algebra has the central terms
computed above.  It is also the setting closest to the conventional
meromorphic OPE derivation of the tensile critical dimension.

The induced null-string vacuum is different.  In the conventional ILST
theory it can eliminate the dimension-selecting central term under a
symmetric regularization \cite{Chen2023}.  Light-cone analyses of the
three standard null-string vacua similarly find that two branches retain
\(D=26\), whereas one places no restriction on \(D\)
\cite{BagchiVacua2021}.  These results reinforce rather than weaken the
need to specify the representation.  A parallel statement for the
complete \(bcs\) system would have to be established directly in the
induced representation; it cannot be inferred by continuing the
highest-weight OPEs.

There is also no immediate contradiction with the reduced light-cone
Schr\"odinger quantization of \cite{Rasulian2026}, which finds \(D-3\)
physical directions and no selected critical dimension.  In that
formalism all three first-class constraints are solved before
quantization.  There is no unreduced \(LMS\) BRST algebra whose central
extensions must cancel.  Equality of the two quantizations would require
a nontrivial equivalence between their Hilbert spaces, measures, and
operator orderings.  The comparison should therefore be phrased as
follows,
\begin{itemize}[leftmargin=2em]
\item the reduced light-cone construction is consistent without
selecting \(D\) within its chosen representation;
\item the minimal covariant highest-weight BRST complex is anomalous for
every \(D\);
\item establishing equivalence, or identifying the precise obstruction
to equivalence, is a separate quantum problem.
\end{itemize}

\subsection*{Possible ways beyond the minimal flat complex}

The no-common-\(D\) result is robust under zero-mode shifts and current
rescalings, but it also indicates what extra structure would be needed.
If an additional matter sector contributes
\((\Delta\cL,\Delta\cS,\Delta\cSS)\), nilpotency requires
\begin{equation}
 \Delta\cL=54-2D,\qquad
 \Delta\cS=D-6,\qquad
 \Delta\cSS=D-4.
\label{eq:extra-matter-condition}
\end{equation}
These are three conditions, not a single central-charge balance.
Non-minimal BRST quartets can change intermediate currents while leaving
cohomology invariant, but only sectors with nontrivial anomaly data can
alter eq.~\eqref{eq:extra-matter-condition}.  Supersymmetric completions are
another natural possibility: matter fermions, superghosts, and a
Carroll-Weyl superpartner would all contribute.  Finally, a genuinely
non-chiral Carrollian operator algebra may organize the anomaly problem
differently from the fixed-time highest-weight realization.

A useful immediate consequence of eq.~\eqref{eq:extra-matter-condition} is
that an ordinary neutral spectator CFT cannot solve the problem.  Such a
sector can change \(\Delta\cL\), but if it has no Carroll-Weyl current
then \(\Delta\cS=\Delta\cSS=0\).  It would require both \(D=6\) and
\(D=4\).  Any viable compensator must therefore contain a weight-one
current \(J\) with
\begin{equation}
 T_{\mathrm{extra}}(z)J(w)
 \sim\frac{D-6}{(z-w)^3}
 +\frac{J(w)}{(z-w)^2}
 +\frac{\partial J(w)}{z-w},
\qquad
 J(z)J(w)\sim\frac{D-4}{(z-w)^2},
\label{eq:compensator-current}
\end{equation}
as well as total Virasoro charge \(54-2D\), in the normalization where
the extra current is added directly to \(S\).
Eq.~\eqref{eq:compensator-current} is a concrete target for a linear-dilaton,
supersymmetric, or other non-minimal construction.

An improvement \(T_{\mathrm{extra}}\to
T_{\mathrm{extra}}+\alpha\partial J\) shifts its mixed anomaly by a
multiple of the affine level, so the three numbers are correlated rather
than freely adjustable.  A proposed completion must also preserve the
\(S M\) charge of \(M\), the vanishing \(LM\) cocycle, and the
nilpotency of any added ghost sector.  Merely matching the first entry
of eq.~\eqref{eq:extra-matter-condition} is insufficient.

\subsection*{Implications of the BRST consistency conditions}

The calculation proves a negative statement with a sharply defined
domain: no value of \(D\) makes the minimal flat highest-weight
\((X,P)\oplus(b,c,\widetilde b,\widetilde c,r,s)\) BRST charge
nilpotent.  It does not prove that the classical Carroll-Weyl string is
inconsistent, that its reduced quantization is empty, or that every
choice of vacuum has the same anomaly.  It also does not turn \(D=27\)
into a physical critical dimension; that number cancels only the
Virasoro cocycle.

There are three logically distinct ways forward.  One may change the
representation and recompute all contractions, enlarge the worldsheet
matter and ghost system so that eq.~\eqref{eq:extra-matter-condition} holds,
or solve the constraints before quantization and then compare the
reduced Hilbert space with a covariant construction.  Each route changes
a different input.  Keeping these alternatives separate prevents a
vacuum-dependent absence of an anomaly from being mistaken for a
cancellation inside the present highest-weight complex.

\section{Conclusions and outlook}
\label{sec:conclusion}

Starting from the Carroll-Weyl gauged action, we have followed the full
quantization chain without inserting a current-algebra realization by
hand.  The canonical moment maps give the matter currents
\((T^X,M^X,S^X)\); the three-row Faddeev-Popov action fixes the ghost
symplectic form, weights, and stress tensor; and the semidirect
structure constants determine the mixed ghost currents.  The cylinder
coupling \(-2sb_0\), the BRST term
\(2s\widetilde b\widetilde c\), and the algebraic relation
\([S,M]=2M\) are consequently three manifestations of the same gauge
structure.

The explicit OPE calculation yields
\[
 (\cL,\cS,\cSS)_X=(2D,-D,-D),\qquad
 (\cL,\cS,\cSS)_\gh=(-54,6,4).
\]
The \(S_0\) Jacobi identity eliminates an independent \(LM\) cocycle,
so these are exactly the three central channels that can enter the
BRST square.  Because they multiply linearly independent ghost
bilinears, nilpotency requires
\begin{equation*}
 2D-54=0,\qquad 6-D=0,\qquad 4-D=0.
\end{equation*}
Their separate zeros, \(D=27,6,4\), are not three candidate critical
dimensions: they are the zeros of three different anomalies.  Since no
single value cancels all of them, \(Q_B\) does not define a cohomology
for the minimal flat flipped-vacuum complex.

Removing the Carroll-Weyl constraint and its scalar ghost reduces the
calculation to the two-constraint BMS subsector and restores the old
\(D=26\) result.  This reduction also deletes the \(LS\) and \(SS\)
tests, so it is not a quantization of the gauge-complete action. We therefore conclude that the minimal Carroll-Weyl BRST charge fails to be nilpotent in the highest-weight representation for every target-space dimension.

The calculation also illustrates three layers of consistency that
should be kept separate in future work.  Classical closure determines
the constraint algebra and the number of ghosts.  Quantum closure
determines which central extensions are allowed.  BRST nilpotency then
requires the total coefficient of every allowed nontrivial cocycle to
vanish.  Passing the first step does not guarantee the second, and
canceling the Virasoro cocycle does not guarantee the third.  This
hierarchy is particularly important for degenerate worldsheets, where a
symmetry missed at the classical level changes the later two steps.
We conclude by outlining some future directions.

The complete \(bcs\) algebra should be reconstructed in the induced
vacuum.  All mixed contractions must follow from the induced
annihilation conditions; importing the flipped-vacuum ghost currents
and then deleting their central terms would not define the same quantum
theory.

The conditional equations
eqs.~\eqref{eq:extra-matter-condition} and \eqref{eq:compensator-current} provide a
systematic target for additional matter.  A neutral spectator is
insufficient: a viable sector must carry a Carroll-Weyl current with
the required background charge and affine level, while preserving the
semidirect action on \(M\).

If a nilpotent completion is found, its relative cohomology, zero-mode
measure, and vertex operators must implement all three constraints.
In particular, the \(S_n\) conditions and the constant \(s\)-ghost mode
modify the homogeneity and ghost-number rules inherited from the
two-constraint BMS string.

Supersymmetric completions are especially natural because matter
fermions, superghosts, and a Carroll-Weyl superpartner modify all three
anomaly coefficients in a correlated way.  Curved backgrounds and
linear-dilaton-like improvements provide another possible route. In the Poincar\'e patch of $\mathrm{AdS}_D$, the choice of Carroll-Weyl
generator changes the anomaly conditions. The flat-space constraint
$X\cdot P$ generates dilatations, whereas the AdS constraint $P_z$
generates radial translations. In the canonical highest-weight
representation of the radial theory, only the Virasoro anomaly
survives, and its coefficient $c_{LL}=2D-54$ vanishes at
$D=27$~\cite{DuaryMajiADS}. The radial coordinate, its conjugate
momentum, and the corresponding ghosts form a contractible quartet
in the BRST complex of local spatial jets. Removing this quartet
leaves the $(D-1)$-dimensional ILST null-string complex. Since the
quartet carries zero net Virasoro central charge, the parent condition
$D=27$ descends to the ILST condition $d=26$%
~\cite{DuaryMajiNullStringHolography}.

It would also be interesting to compare our findings with the
complementary spacetime perspective of
\cite{Ballesteros:2026bqe,Lescano:2026hxd}, where the Carrollian limit is studied directly from the supergravity side. There,
both the bosonic and the heterotic theories are found to be finite
independently of the spacetime dimension: their finiteness
concerns the target-space effective action, whereas the obstruction we
find here arises from BRST nilpotency of the minimal Carroll-Weyl
worldsheet theory. Clarifying how these two notions of consistency are
related -- in particular, whether the dimension-independent finiteness
of the effective action survives, or is reinterpreted, once the
worldsheet constraints derived above are imposed -- would sharpen the
connection between tensionless strings and their effective target-space
descriptions.

\section*{Acknowledgments}

We thank Zezhou Hu for useful discussions and valuable comments on the draft. We also thank Bin Chen and Zezhou Hu for sharing their draft with us and coordinating the release of our papers. We thank Eric Lescano for valuable comments on the draft. SM thanks Adarsh S for useful discussions. SD is supported by the Shuimu Tsinghua Scholar Program of Tsinghua University and the Beijing Natural Science Foundation of China under Grant No.~IS25035. 

\appendix

\section{Cylinder modes and the equal-time OPE dictionary}
\label{app:cylinder}

\subsection*{Matter equations, modes, and symplectic form}

In the phase-space gauge \(e=1,u=0\), the null action
eq.~\eqref{eq:null-phase-space} becomes
\begin{equation}
 S_X=\frac{1}{2\pi}\int\dd\tau\,\dd\sigma
 \left(P\cdot\dot X-\frac12P^2\right).
\label{eq:app-phase-action}
\end{equation}
The equations of motion and their general closed-string solution are
\begin{equation}
 \dot X^\mu=P^\mu,\qquad \dot P^\mu=0,
\qquad
 X^\mu(\tau,\sigma)=x^\mu(\sigma)+\tau p^\mu(\sigma),
\qquad P^\mu(\tau,\sigma)=p^\mu(\sigma).
\label{eq:matter-cylinder-solution}
\end{equation}
Writing
\begin{equation}
 x^\mu(\sigma)=\sum_{n\in\mathbb Z}x^\mu_n\e^{-\ii n\sigma},
 \qquad
 p_\mu(\sigma)=\sum_{n\in\mathbb Z}p_{\mu,n}\e^{-\ii n\sigma},
\label{eq:matter-Fourier}
\end{equation}
the symplectic form at a reference slice is
\begin{equation}
 \Omega_X
 =\frac{1}{2\pi}\int_0^{2\pi}\dd\sigma\,
 \delta P_\mu\wedge\delta X^\mu
 =\sum_{n\in\mathbb Z}
 \delta p_{\mu,-n}\wedge\delta x^\mu_n.
\label{eq:matter-symplectic}
\end{equation}
Quantization gives
\([x^\mu_m,p_{\nu,n}]=\ii\delta^\mu{}_\nu\delta_{m+n,0}\);
the factor of \(\ii\) is absorbed into the Laurent convention used in
eq.~\eqref{eq:XP-mode-bracket}.  Zero modes obey the same canonical algebra
but require their own choice of wavefunction representation.

At \(\tau=0\), tensor-density factors map the cylinder variables to the
weight-\((0,1)\) plane fields:
\begin{equation}
 X^\mu(z)=x^\mu(\sigma),\qquad
 P_\mu(z)=z^{-1}p_\mu(\sigma),\qquad
 z=\e^{\ii\sigma}.
\label{eq:matter-cylinder-plane}
\end{equation}
The factor \(z^{-1}\) is precisely what turns \(p_\mu\) into a
weight-one Laurent field.  The unequal Carrollian-time solution is
recovered from
\(X(\tau,\sigma)=X(0,\sigma)+\tau P(0,\sigma)\).

\subsection*{Constraint brackets in detail}

Use the smeared generators in eq.~\eqref{eq:smeared-MLSa}.  Functional
differentiation gives
\begin{align}
 \frac{\delta C_1[f]}{\delta P_\mu}&=2fP^\mu,
 &
 \frac{\delta C_2[g]}{\delta P_\mu}&=gX'^\mu,
 &
 \frac{\delta C_2[g]}{\delta X^\mu}&=-(gP_\mu)',
\nonumber\\
 \frac{\delta C_3[h]}{\delta P_\mu}&=hX^\mu,
 &
 \frac{\delta C_3[h]}{\delta X^\mu}&=hP_\mu .
\label{eq:functional-derivatives}
\end{align}
For example,
\begin{align}
 \{C_3[h],C_1[f]\}
 &=
 \int\dd\sigma\,
 \frac{\delta C_3[h]}{\delta X^\mu}
 \frac{\delta C_1[f]}{\delta P_\mu}
 =2\int\dd\sigma\,hfP^2
 =2C_1[hf],
\label{eq:derive-SM}\\
 \{C_2[g],C_3[h]\}
 &=
 \int\dd\sigma\left[
 -(gP_\mu)'hX^\mu-gX'^\mu hP_\mu
 \right]
 =C_3[gh'] ,
\label{eq:derive-LS}
\end{align}
where the circle has no boundary.  Similarly,
\begin{align}
 \{C_1[f],C_2[g]\}&=C_1[fg'-f'g],&
 \{C_2[g],C_2[k]\}&=C_2[gk'-g'k],\nonumber\\
 \{C_1[f],C_3[h]\}&=-2C_1[fh],&
 \{C_3[h],C_3[\ell]\}&=0,\qquad
 \{C_1[f],C_1[q]\}=0.
\label{eq:derived-smeared-algebra}
\end{align}
These are the brackets in the exact convention
\((C_1,C_2,C_3)=(P^2,P\cdot X',P\cdot X)\).  Applying the dictionary
in eq.~\eqref{eq:constraint-to-LMS} and Fourier expanding gives
eq.~\eqref{eq:centerless-LMS}.  Eqs.~\eqref{eq:derive-SM} and
\eqref{eq:derive-LS} also fix the weights and charges used to construct
the ghosts.

\subsection*{Linearized gauge transformations}

For completeness, let \(\epsilon^a\) generate diffeomorphisms and
\(\lambda\) the infinitesimal Carroll-Weyl rescaling.  Since \(V^a\)
is a vector density of weight \(1/2\),
\begin{equation}
 \delta V^a
 =-\epsilon^b\partial_bV^a+V^b\partial_b\epsilon^a
 -\frac12V^a\partial_b\epsilon^b-\lambda V^a.
\label{eq:V-transformation-app}
\end{equation}
The contraction \(W=V^aW_a\) transforms, to linear order about
\(V^a=(1,0),W=0\), as
\begin{equation}
 \delta W=-\partial_\tau\lambda.
\label{eq:W-transformation-app}
\end{equation}
Evaluating eq.~\eqref{eq:V-transformation-app} on the same slice yields
\begin{align}
 \delta(V^0-1)&=
 \frac12\partial_\tau\epsilon^0
 -\frac12\partial_\sigma\epsilon^1-\lambda,
\nonumber\\
 \delta V^1&=\partial_\tau\epsilon^1.
\label{eq:V-linear-app}
\end{align}
Changing all diffeomorphism parameters by an overall sign gives the
convention in eq.~\eqref{eq:gauge-variations}; the determinant and ghost
system are unchanged.  Ordering the gauge functions and parameters
then gives the triangular operator eq.~\eqref{eq:FP-operator}.

The formal determinant factorizes into three time derivatives, but the
matrix itself does not.  Exponentiation retains the upper-right entry
\begin{align}
 -b_A(\mathcal M_{\CW})^A{}_Bc^B
 ={}&
 b_0\partial_\tau c^0-b_0\partial_\sigma c^1+2b_0s
\nonumber\\
 &+2b_1\partial_\tau c^1+b_s\partial_\tau s .
\label{eq:FP-before-parts}
\end{align}
Grassmann integration by parts gives eq.~\eqref{eq:bcs-action}.  Dropping
the off-diagonal term because it does not change the numerical
determinant would lose the correct BRST transformations and is not a
legitimate diagonalization of the gauge complex.

\subsection*{Solutions of the ghost equations}

On the closed-string cylinder, \(\sigma\sim\sigma+2\pi\), the solutions
of eq.~\eqref{eq:ghost-eom} are
\begin{align}
 c^1(\tau,\sigma)
 &=\sum_{n\in\mathbb Z}c_n\e^{-\ii n\sigma},
 &
 s(\tau,\sigma)
 &=\sum_{n\in\mathbb Z}s_n\e^{-\ii n\sigma},
\label{eq:cyl-c-s}\\
 c^0(\tau,\sigma)
 &=\sum_{n\in\mathbb Z}
 \left(\widetilde c_n-\ii n\tau c_n-2\tau s_n\right)
 \e^{-\ii n\sigma},
\label{eq:cyl-c0}\\
 b_0(\tau,\sigma)
 &=\sum_{n\in\mathbb Z}\widetilde b_n\e^{-\ii n\sigma},
 &
 b_1(\tau,\sigma)
 &=\half\sum_{n\in\mathbb Z}
 \left(b_n-\ii n\tau\widetilde b_n\right)
 \e^{-\ii n\sigma},
\label{eq:cyl-b}\\
 b_s(\tau,\sigma)
 &=\sum_{n\in\mathbb Z}
 \left(r_n+2\tau\widetilde b_n\right)\e^{-\ii n\sigma}.
\label{eq:cyl-bs}
\end{align}
The terms \(-2\tau s_n\) and \(2\tau\widetilde b_n\) are generated by
the off-diagonal Faddeev-Popov coupling.

At \(\tau=0\), the symplectic one-form following from
eq.~\eqref{eq:bcs-action} is
\begin{equation}
 \Theta_{bcs}
 =\frac{\ii}{2\pi}\int_0^{2\pi}\dd\sigma\,
 \left(c^0\delta b_0+2c^1\delta b_1+s\delta b_s\right)
 =\ii\sum_n\left(
 \widetilde c_{-n}\delta\widetilde b_n
 +c_{-n}\delta b_n+s_{-n}\delta r_n
 \right).
\label{eq:ghost-symplectic}
\end{equation}
It gives eq.~\eqref{eq:ghost-oscillators}.  The algebraic term
\(-2sb_0\) changes Hamiltonian evolution but not the equal-time
symplectic form.

\subsection*{From Fourier modes to Laurent fields}

A field of conformal weight \(h\) has the Laurent expansion
\begin{equation}
 \Phi(z)=\sum_{n\in\mathbb Z}\Phi_n z^{-n-h}.
\label{eq:weight-Laurent}
\end{equation}
The six reference-slice fields are packaged as
\begin{align}
 b(z)&=\sum_n b_nz^{-n-2},
 &
 c(z)&=\sum_n c_nz^{-n+1},
\nonumber\\
 \widetilde b(z)&=\sum_n\widetilde b_nz^{-n-2},
 &
 \widetilde c(z)&=\sum_n\widetilde c_nz^{-n+1},
\nonumber\\
 r(z)&=\sum_n r_nz^{-n-1},
 &
 s(z)&=\sum_n s_nz^{-n}.
\label{eq:ghost-Laurent}
\end{align}
Since \(z=\e^{\ii\sigma}\), the explicit dictionary at \(\tau=0\) is
\begin{equation}
 c(z)=z\,c^1,\quad
 b(z)=2z^{-2}b_1,\quad
 \widetilde c(z)=z\,c^0,\quad
 \widetilde b(z)=z^{-2}b_0,\quad
 s(z)=s,\quad
 r(z)=z^{-1}b_s .
\label{eq:explicit-dictionary}
\end{equation}
The extra powers of \(z\) are tensor-density factors; the Laurent fields
are not new solutions of the cylinder equations.

Choosing a highest-weight vacuum splits every Laurent field into
annihilation and creation parts.  For \(|z|>|w|\),
\begin{equation}
 \big(b(z)c(w)\big)_{\mathrm{sing}}
 =\frac{1}{z-w},
\label{eq:bc-from-modes}
\end{equation}
and similarly for the other two pairs.  Thus the logical chain is
\[
\begin{gathered}
 \text{cylinder action}
 \ \Longrightarrow\
 \text{canonical oscillators}
 \ \Longrightarrow\
 \text{conformal weights and highest-weight splitting}
 \\
 \Longrightarrow\ \text{equal-time OPEs}.
\end{gathered}
\]

\subsection*{A useful Wick-contraction check}

For the fermionic pair in eq.~\eqref{eq:first-order-stress}, set
\[
 A=\normal{(\partial B)C},\qquad
 B_1=\normal{B\partial C},\qquad
 T_{BC}=(1-h)A-hB_1.
\]
The double contractions are
\begin{equation}
 A(z)A(w)\big|_{\mathrm{double}}
 =-\frac{1}{(z-w)^4},\quad
 B_1(z)B_1(w)\big|_{\mathrm{double}}
 =-\frac{1}{(z-w)^4},
\end{equation}
and
\begin{equation}
 \big[A(z)B_1(w)+B_1(z)A(w)\big]_{\mathrm{double}}
 =-\frac{4}{(z-w)^4}.
\end{equation}
The coefficient of the fourth-order pole is
\[
 -(1-h)^2-h^2+4h(1-h)=-6h^2+6h-1,
\]
which is \(c(h)/2\) and proves eq.~\eqref{eq:bc-central-charge}.

\section{Ghost currents from the BFV differential}
\label{app:ghost}
The main text derives the currents directly from the gauge-fixed action and presents all contractions required to determine the anomaly vector.  This appendix complements that analysis by establishing the normalization dictionary and providing a sign-complete BFV consistency check.  Its purpose is to make the semidirect ghost currents and their algebra independently reproducible while keeping the main discussion focused on the central result.


\subsection*{Mapping from unit-charge Weyl-BMS variables}

It is useful to begin with generators
\((T,\mathcal D,M)\) satisfying
\begin{equation}
 [\mathcal D_m,M_n]=M_{m+n}.
\end{equation}
Let their ghosts be \((c,c_{\mathcal D},\widetilde c)\) and antighosts
\((b,b_{\mathcal D},\widetilde b)\).  The physical null-string
normalization is
\begin{equation}
 S=2\mathcal D,\qquad
 c_{\mathcal D}=2s,\qquad
 b_{\mathcal D}=\half r.
\label{eq:ghost-rescale-map}
\end{equation}
The rescaling preserves
\(b_{\mathcal D}(z)c_{\mathcal D}(w)\sim1/(z-w)\) and the term
\(c_{\mathcal D}\mathcal D=sS\) in the BRST current.

For the \(\lambda=-1\) Weyl-BMS algebra, the unit-charge ghost currents
are \cite{Batlle2024}
\begin{align}
 \mathcal D^\gh
 &=\normal{\widetilde b\widetilde c}
 +\partial\normal{c b_{\mathcal D}},
\label{eq:Dgh-unit}\\
 M^\gh
 &=\normal{c\partial\widetilde b}
 +\normal{c_{\mathcal D}\widetilde b}
 +2\normal{(\partial c)\widetilde b}.
\label{eq:Mgh-unit}
\end{align}
Substituting eq.~\eqref{eq:ghost-rescale-map} and using
\(S^\gh=2\mathcal D^\gh\) gives
eqs.~\eqref{eq:Sghost} and \eqref{eq:Mghost}.

\subsection*{BRST transformations}

The local BFV current is
\begin{align}
 j_B={}&
 \normal{cT^X}+\normal{sS^X}+\normal{\widetilde cM^X}
 +\normal{bc\partial c}+\normal{rc\partial s}
\nonumber\\
 &+\normal{\widetilde b c\partial\widetilde c}
 +\normal{\widetilde b\widetilde c\partial c}
 +2\normal{s\widetilde b\widetilde c}.
\end{align}
Contracting with the ghost fields gives
\begin{equation}
 \delta_Bc=c\partial c,\qquad
 \delta_Bs=c\partial s,\qquad
 \delta_B\widetilde c
 =c\partial\widetilde c+\widetilde c\partial c
 +2s\widetilde c.
\end{equation}
Contracting with the antighosts gives the total constraints in
eq.~\eqref{eq:Q-antighost}.  In particular, the last term in \(j_B\) is
simultaneously required by \(\delta_B\widetilde c\), by
\([S,M]=2M\), and by the Faddeev-Popov mixing.

\subsection*{BRST transformations of matter and antighosts}

Using eqs.~\eqref{eq:T-on-XP} and \eqref{eq:S-on-XP}, the matter transformations generated
by the contour charge are
\begin{align}
 \delta_BX^\mu
 &=c\,\partial X^\mu-\widetilde c\,P^\mu-sX^\mu,
\label{eq:BRST-X}\\
 \delta_BP_\mu
 &=c\,\partial P_\mu+(\partial c)P_\mu+sP_\mu.
\label{eq:BRST-P}
\end{align}
The relative signs follow from the radial convention in
eq.~\eqref{eq:matter-basic-ope}.  The three terms in
eq.~\eqref{eq:BRST-X} are, respectively, the spatial reparametrization,
null translation along \(P^\mu\), and target homothety induced by the
Carroll-Weyl constraint.  Combining
eqs.~\eqref{eq:BRST-X} and \eqref{eq:BRST-P} with the ghost transformations gives
\(\delta_B^2X=\delta_B^2P=0\) at the classical, centerless level.

The antighost transformations are more informative quantum mechanically:
\begin{align}
 \delta_Bb&=T^X+T^\gh,&
 \delta_B\widetilde b&=M^X+M^\gh,&
 \delta_Br&=S^X+S^\gh.
\label{eq:BRST-antighosts-app}
\end{align}
For instance, contracting \(r(z)\) with
\(2\normal{s\widetilde b\widetilde c}(w)\) supplies
\(2\normal{\widetilde b\widetilde c}\), while its contraction with
\(\normal{rc\partial s}\) supplies the
\(\partial\normal{cr}\) part of \(S^\gh\), after a contour integration
by parts.  Thus the two terms of eq.~\eqref{eq:Sghost} are both visible
directly in the BFV current.

\subsection*{Representative ghost OPEs}

For clarity, the non-central terms and the possible central contractions
are summarized in \cref{tab:ghost-ope-detail}.

\begin{table}[H]
\centering
\renewcommand{\arraystretch}{1.2}
\begin{tabular}{|p{0.20\textwidth}|p{0.42\textwidth}|p{0.22\textwidth}|}
\hline
OPE & Non-central singular part & Central part\\
\hline
\(T^\gh T^\gh\)
& \(2T^\gh/(z-w)^2+\partial T^\gh/(z-w)\)
& \(-27/(z-w)^4\)\\
\(T^\gh S^\gh\)
& \(S^\gh/(z-w)^2+\partial S^\gh/(z-w)\)
& \(6/(z-w)^3\)\\
\(S^\gh S^\gh\)
& \(0\) & \(4/(z-w)^2\)\\
\(T^\gh M^\gh\)
& \(2M^\gh/(z-w)^2+\partial M^\gh/(z-w)\)
& \(0\)\\
\(S^\gh M^\gh\)
& \(2M^\gh/(z-w)\) & \(0\)\\
\(M^\gh M^\gh\)
& \(0\) & \(0\)\\
\hline
\end{tabular}
\caption{The equal-time ghost current algebra.}
\label{tab:ghost-ope-detail}
\end{table}

Two quick structural checks are worth recording.  First, every term in
\(M^\gh\) contains \(\widetilde b\) but no \(\widetilde c\); hence
\(M^\gh M^\gh\) cannot have a surviving double contraction.  Second,
the contraction of \(\partial(cr)\) with \(2s\widetilde b\) is necessary
for the derivative terms in \(S^\gh M^\gh\) to recombine into
\(2M^\gh/(z-w)\).  Omitting the scalar ghost pair would destroy this
closure even before central terms are considered.

\subsection*{Double contractions and central terms}

Here we display the contractions responsible for the three ghost
anomalies.  For a weight-\((h,1-h)\) pair, write
\begin{equation}
 T_h=(1-h)\normal{(\partial B)C}
      -h\normal{B\partial C}.
\label{eq:Th-ledger}
\end{equation}
The fourth-order terms are
\begin{align}
 \normal{(\partial B)C}(z)
 \normal{(\partial B)C}(w)\big|_2
 &=-\frac{1}{(z-w)^4},
\nonumber\\
 \normal{B\partial C}(z)
 \normal{B\partial C}(w)\big|_2
 &=-\frac{1}{(z-w)^4},
\nonumber\\
 \left[
 \normal{(\partial B)C}(z)\normal{B\partial C}(w)
 +(z\leftrightarrow w)
 \right]\big|_2
 &=-\frac{4}{(z-w)^4}.
\label{eq:bc-double-ledger}
\end{align}
Including the coefficients in eq.~\eqref{eq:Th-ledger}, the pole is
\((-6h^2+6h-1)/(z-w)^4=c(h)/[2(z-w)^4]\).
For \(h=2,2,1\) the three entries are
\(-13,-13,-1\), adding to \(-27\).

For the mixed anomaly, only the
\((\widetilde b,\widetilde c)\) stress tensor can doubly contract with
\(2N_M=2\normal{\widetilde b\widetilde c}\),
\begin{align}
 &\left[
 -2\normal{\widetilde b\partial\widetilde c}
 -\normal{(\partial\widetilde b)\widetilde c}
 \right](z)\,
 2\normal{\widetilde b\widetilde c}(w)
 \Big|_2
 =\frac{6}{(z-w)^3}.
\label{eq:TSghost-ledger}
\end{align}
The composite \(A=\normal{cr}\) has no contraction with \(N_M\).
The OPE \(T^\gh(z)\partial A(w)\) supplies weight-one descendant terms
but no field-independent third-order pole.

Finally,
\begin{equation}
 4N_M(z)N_M(w)\Big|_2=\frac{4}{(z-w)^2}
\label{eq:SSghost-ledger}
\end{equation}
is the entire affine level.  The products involving
\(\partial A\) contain no double contractions because \(c\) and \(r\)
belong to different canonical pairs.
Eqs.~\eqref{eq:bc-double-ledger}--\eqref{eq:SSghost-ledger} reproduce
\((\cL,\cS,\cSS)_\gh=(-54,6,4)\) without appealing to an abstract
central-charge formula.

\subsection*{Closure of the semidirect current}

We set
\begin{equation}
 M_1=\normal{c\partial\widetilde b},\qquad
 M_2=2\normal{s\widetilde b},\qquad
 M_3=2\normal{(\partial c)\widetilde b}.
\label{eq:Mghost-pieces}
\end{equation}
All three have weight two.  The number-current part
\(2N_M\subset S^\gh\) contracts with the \(\widetilde b\) in every
\(M_i\).  The improvement
\(\partial\normal{cr}\) contracts with the \(s\) in \(M_2\).
After Taylor expanding the surviving fields, the derivative terms from
these two sources combine as
\begin{equation}
 S^\gh(z)\big(M_1+M_2+M_3\big)(w)
 \sim\frac{2(M_1+M_2+M_3)(w)}{z-w}.
\label{eq:SMghost-pieces}
\end{equation}
If either \(M_2\) or the improvement in \(S^\gh\) is dropped, uncancelled
terms proportional to \(\partial c\,\widetilde b\) remain and
eq.~\eqref{eq:SMghost-pieces} fails.  This explicitly locates the
semidirect-product information inside the OPE.

\section{Mode algebra, cocycles, and intercepts}
\label{app:modes}

\subsection*{From local fields to generators}

The mode extraction contours are
\begin{equation}
 L_m=\oint\frac{\dd z}{2\pi\ii}\,z^{m+1}T(z),\qquad
 M_m=\oint\frac{\dd z}{2\pi\ii}\,z^{m+1}M(z),\qquad
 S_m=\oint\frac{\dd z}{2\pi\ii}\,z^mS(z).
\label{eq:mode-contours}
\end{equation}
If \(A(z)B(w)\) is known for \(|z|>|w|\), their graded commutator is
\begin{equation}
 [A_m,B_n]_\pm
 =\oint_0\frac{\dd w}{2\pi\ii}\,w^{n+h_B-1}
 \oint_w\frac{\dd z}{2\pi\ii}\,z^{m+h_A-1}
 A(z)B(w).
\label{eq:double-contour}
\end{equation}
The inner contour isolates the singular OPE.  For example, the
second- and first-order terms in \(T(z)M(w)\) give
\begin{align}
 [L_m,M_n]
 &=
 \oint_0\frac{\dd w}{2\pi\ii}\,w^{n+1}
 \left[
 2(m+1)w^mM(w)+w^{m+1}\partial M(w)
 \right]
\nonumber\\
 &=(m-n)M_{m+n}.
\label{eq:derive-LM-modes}
\end{align}
The integration by parts in the last step is the mode counterpart of
the Taylor expansion in a single contraction.

\subsection*{From OPE poles to mode central terms}

We use
\begin{equation}
 T(z)=\sum_m L_mz^{-m-2},\qquad
 M(z)=\sum_m M_mz^{-m-2},\qquad
 S(z)=\sum_m S_mz^{-m-1}.
\end{equation}
The contour residues needed for the central terms are
\begin{align}
 \Res_{z=w}\frac{z^{m+1}}{(z-w)^4}
 &=\frac{m(m^2-1)}{6}w^{m-2},\\
 \Res_{z=w}\frac{z^{m+1}}{(z-w)^3}
 &=\frac{m(m+1)}{2}w^{m-1},\\
 \Res_{z=w}\frac{z^m}{(z-w)^2}
 &=m\,w^{m-1}.
\end{align}
They give eqs.~\eqref{eq:mode-LL}, \eqref{eq:mode-LS}, and \eqref{eq:mode-SS}.

\subsection*{Jacobi classification of the central polynomials}

Suppose the non-central brackets are fixed as in
eq.~\eqref{eq:centerless-LMS} and write the possible central terms as
\begin{align}
 [L_m,L_n]_{\mathrm{cen}}
 &=A(m)\delta_{m+n,0},
\nonumber\\
 [L_m,S_n]_{\mathrm{cen}}
 &=B(m)\delta_{m+n,0},
\nonumber\\
 [S_m,S_n]_{\mathrm{cen}}
 &=K(m)\delta_{m+n,0}.
\label{eq:ABC-central}
\end{align}
Antisymmetry gives \(A(-m)=-A(m)\) and \(K(-m)=-K(m)\).  The
\((L,L,L)\) Jacobi identity gives the familiar recurrence
\begin{equation}
 (m-n)A(m+n)-(m+2n)A(m)+(2m+n)A(n)=0,
\label{eq:Virasoro-recurrence}
\end{equation}
whose local polynomial solution is
\begin{equation}
 A(m)=a_3m^3+a_1m.
\label{eq:A-solution}
\end{equation}
The \(a_1\) term is a coboundary shifted by \(L_0\); choosing the
standard global-\(\mathfrak{sl}(2)\) representative gives
\(A(m)=\cL m(m^2-1)/12\).

The \((L,L,S)\) identity similarly fixes, up to a shift of \(S_0\),
\begin{equation}
 B(m)=\frac{\cS}{2}m(m+1),
\label{eq:B-solution}
\end{equation}
while the \((L,S,S)\) identity fixes
\begin{equation}
 K(m)=\cSS m.
\label{eq:K-solution}
\end{equation}
These are three independent cohomology classes.  Their different
polynomial degrees are reflected in the fourth-, third-, and
second-order poles of eq.~\eqref{eq:central-ope-conventions}.

\subsection*{The \texorpdfstring{\(LM\)}{LM} Jacobi identity}

Suppose temporarily that
\begin{equation}
 [L_m,M_n]=(m-n)M_{m+n}
 +K_{LM}(m)\delta_{m+n,0}.
\end{equation}
The Jacobi identity
\[
 [S_0,[L_m,M_n]]+[L_m,[M_n,S_0]]
 +[M_n,[S_0,L_m]]=0
\]
uses \([S_0,M_n]=2M_n\) and \([S_0,L_m]=0\).  The non-central
terms cancel and the remainder is
\[
 -2K_{LM}(m)\delta_{m+n,0}=0.
\]
Thus the BMS \(LM\) cocycle does not extend to the
Carroll-Weyl algebra.

The same conclusion follows from the \((S_p,L_m,M_n)\) Jacobi identity
at arbitrary \(p\), but \(p=0\) is already sufficient because \(S_0\)
measures a nonzero charge of \(M_n\).  In cohomological language, the
usual BMS \(LM\) cocycle is not invariant under adjoining the
dilatation derivation.

\subsection*{Normal-ordering shifts}

The most general Virasoro central polynomial may be written
\[
 A_{LL}(m)=a_3m^3+a_1m.
\]
A shift of \(L_0\) changes \(a_1\) but not \(a_3\).  In the present
realization,
\[
 a_3=\frac{D-27}{6}.
\]
Similarly, the nontrivial coefficients in the scaling sector are
proportional, up to convention-dependent factors, to
\[
 a_{SS}\propto\frac{D}{4}-1,\qquad
 a_{LS}\propto\frac{D}{4}-\frac32.
\]
Their zeros are \(D=4\) and \(D=6\), matching the OPE calculation.
A shift of \(S_0\) changes only a linear coboundary in the mixed
commutator.  The independent zeros \(27,4,6\) therefore cannot be
reconciled by normal ordering.

\subsection*{Mode form of the BRST charge}

Writing \(T^A=(L,M,S)\), \(c^A=(c,\widetilde c,s)\), and
\(b_A=(b,\widetilde b,r)\), the minimal mode charge has the BFV form
\begin{equation}
 Q_B=\sum_{n,A}c^A_{-n}T^A_n
 -\half\sum_{m,n}f_{AB}{}^C(m,n)
 \normal{c^A_{-m}c^B_{-n}b_{C,m+n}}
 +Q_{\mathrm{intercept}}.
\end{equation}
The central part of its square is eq.~\eqref{eq:Q2-central}.  The \(cc\),
\(cs\), and \(ss\) oscillator bilinears are linearly independent.
This is the mode-algebra reason that all three cocycles, rather than a
single combination, must vanish.

For reference, expanding the local representative
eq.~\eqref{eq:BRST-current-explicit} gives the matter part
\begin{equation}
 Q_{\mat}
 =\sum_n\left(
 c_{-n}L^X_n+\widetilde c_{-n}M^X_n+s_{-n}S^X_n
 \right)
\label{eq:Qmatter-modes}
\end{equation}
and cubic terms of the schematic form
\begin{align}
 Q_{\gh}={}&
 -\frac12\sum_{m,n}(m-n)
 \normal{c_{-m}c_{-n}b_{m+n}}
\nonumber\\
 &+\sum_{m,n}n\,
 \normal{c_{-m}s_{-n}r_{m+n}}
\nonumber\\
 &+\sum_{m,n}(n-m)
 \normal{c_{-m}\widetilde c_{-n}\widetilde b_{m+n}}
\nonumber\\
 &-2\sum_{m,n}
 \normal{s_{-m}\widetilde c_{-n}\widetilde b_{m+n}},
\label{eq:Qghost-modes}
\end{align}
where simultaneous sign reversals can result from changing the
definition of all Fourier generators.  The relative coefficient in the
last line is invariant and encodes \([S,M]=2M\).  Contracting
eq.~\eqref{eq:Qghost-modes} with the ghost modes reproduces
eqs.~\eqref{eq:brst-c}, \eqref{eq:brst-s}, and \eqref{eq:brst-ctilde}.

\subsection*{Central part of the BRST square}

Separate the charge into the part linear in the total constraints and
the remaining cubic ghost terms,
\begin{equation}
 Q_{\mathrm{lin}}
 =\sum_n\left(
 c_{-n}L_n^{\tot}
 +\widetilde c_{-n}M_n^{\tot}
 +s_{-n}S_n^{\tot}
 \right),\qquad
 Q_B=Q_{\mathrm{lin}}+Q_{\mathrm{cub}}.
\label{eq:Q-linear-cubic}
\end{equation}
The non-central part of \(Q_{\mathrm{lin}}^2\) is cancelled by
\(\{Q_{\mathrm{lin}},Q_{\mathrm{cub}}\}+Q_{\mathrm{cub}}^2\).
This is the operator form of classical BFV nilpotency.  A central term
commutes with all antighosts and constraints, so no cubic contribution
can remove it.

Using eqs.~\eqref{eq:mode-LL}, \eqref{eq:mode-LS}, and \eqref{eq:mode-SS}, the Virasoro contribution
is
\begin{align}
 \frac12\sum_{m,n}c_{-m}c_{-n}[L_m,L_n]_{\mathrm{cen}}
 &=
 \frac{\cL^\tot}{24}
 \sum_m m(m^2-1)c_{-m}c_m.
\label{eq:Q2-LL-derive}
\end{align}
The two orderings of the \(LS\) bracket give
\begin{align}
 Q_B^2\big|_{LS}
 &=
 \frac{\cS^\tot}{4}
 \sum_m m(m+1)
 \left(c_{-m}s_m-s_{-m}c_m\right),
\label{eq:Q2-LS-derive}
\end{align}
where a change of cocycle representative shifts only the linear
zero-mode part.  Finally,
\begin{align}
 \frac12\sum_{m,n}s_{-m}s_{-n}[S_m,S_n]_{\mathrm{cen}}
 &=
 \frac{\cSS^\tot}{2}\sum_m m\,s_{-m}s_m.
\label{eq:Q2-SS-derive}
\end{align}
The \(LM\) term is absent by eq.~\eqref{eq:LM-Jacobi}; the \(SM\) and \(MM\)
brackets carry no central extension.

Independence can be checked without invoking abstract cohomology.
Matrix elements between ghost Fock states containing only two
\(c\)-oscillators project onto eq.~\eqref{eq:Q2-LL-derive}.  States with
one \(c\) and one \(s\) project onto eq.~\eqref{eq:Q2-LS-derive}, and
states with two \(s\)-oscillators project onto
eq.~\eqref{eq:Q2-SS-derive}.  Since
each matrix element can be nonzero while the other two vanish, a
cancellation among the three coefficients is impossible.

\subsection*{Anomaly consistency and counterterms}

The central terms satisfy the Wess-Zumino consistency condition because
they are Lie-algebra two-cocycles.  A local redefinition of a current
adds a coboundary.  In modes, shifts of \(L_0\) and \(S_0\) alter the
linear pieces of \(A(m)\) and \(B(m)\); in OPE language, this changes
contact terms or total derivatives.  The coefficients of
\(m^3\), \(m^2\), and \(m\) in
eqs.~\eqref{eq:Q2-LL-derive}--\eqref{eq:Q2-SS-derive} are invariant under
such changes.

This separates regularization dependence from counterterm freedom.
Choosing a different vacuum can change the contraction itself and
hence the cocycle coefficients.  Once the highest-weight contraction
and point-splitting prescription are fixed, however, local counterterms
cannot turn the vector
\((2D-54,6-D,4-D)\) into zero unless all three entries already vanish.

\subsection*{Zero modes and ghost-number bookkeeping}

On the auxiliary spatial plane, each weight-\((-1,2)\) ghost pair has
the three global Witt modes customarily associated with
\(n=-1,0,1\), while the weight-\((0,1)\) \(s,r\) pair has a constant
ghost mode.  The minimal ghost-number current may be chosen as
\begin{equation}
 j_{\gh}
 =-\normal{bc}
  -\normal{\widetilde b\widetilde c}
  -\normal{rs},
\label{eq:ghost-number-current}
\end{equation}
up to an overall sign convention.  Its contour assigns ghost number
\(+1\) to \(c,\widetilde c,s\) and \(-1\) to their antighosts.

The zero-mode part \(Q_{\mathrm{intercept}}\) in
eqs.~\eqref{eq:Qmatter-modes} and \eqref{eq:Qghost-modes} imposes the analogues of the
mass-shell and intercept conditions.  These affect the linear
coboundary pieces of the mode algebra and the ghost-number selection
rule for amplitudes.  They do not touch the \(m^3\), \(m^2\), or
affine-\(m\) coefficients responsible for the three conditions in
eq.~\eqref{eq:three-D-equations}.


\section{OPE characterization of the BRST current as a tensor}
\label{app:BRST-current-transformation}

For the tensile bosonic string, the critical-dimension condition may be
recovered by requiring the BRST current to transform as a conformal field of
weight one. This appendix develops the corresponding criterion for the
Carroll-Weyl gauged null string. The extension is nontrivial because the
quantum constraint algebra contains three independent cocycles, $c_{LL}^{\mathrm{tot}},
c_{LS}^{\mathrm{tot}},
c_{SS}^{\mathrm{tot}}$.
Accordingly, the transformation of the BRST current must be examined under
both \(T^{\mathrm{tot}}\) and 
\(S^{\mathrm{tot}}\). The first transformation detects the \((LL)\) and
\((LS)\) anomalies, whereas the second also detects the affine \((SS)\)
anomaly.

\subsection*{OPEs with $T^{\mathrm{tot}}$, $S^{\mathrm{tot}}$, $M^{\mathrm{tot}}$}
We now calculate the OPEs of BRST current $j_B$ with $T^{\mathrm{tot}}$, $S^{\mathrm{tot}}$, $M^{\mathrm{tot}}$.
\subsubsection*{Transformation under $T^{\mathrm{tot}}$.}

The noncentral contributions to
\(\bigl[L_m^{\mathrm{tot}},Q_B\bigr]\) cancel against the cubic ghost terms
by the BFV construction. Its anomalous part therefore comes entirely from
the \((LL)\) and \((LS)\) central extensions
\begin{align}
\bigl[
L_m^{\mathrm{tot}},
Q_B
\bigr]_{\mathrm{anom}}
={}&
\sum_n c_{-n}
\bigl[
L_m^{\mathrm{tot}},
L_n^{\mathrm{tot}}
\bigr]_{\mathrm{cen}}
+
\sum_n s_{-n}
\bigl[
L_m^{\mathrm{tot}},
S_n^{\mathrm{tot}}
\bigr]_{\mathrm{cen}}
\nonumber\\
={}&
\frac{c_{LL}^{\mathrm{tot}}}{12}
m(m^2-1)c_m
+
\frac{c_{LS}^{\mathrm{tot}}}{2}
m(m+1)s_m.
\label{eq:app-Lm-QB-anomaly}
\end{align}

To reconstruct the corresponding local OPE, consider
\begin{equation}
T^{\mathrm{tot}}(z)j_B(w)
\sim
\frac{A\,c(w)}{(z-w)^4}
+
\frac{B\,s(w)}{(z-w)^3}
+
\frac{j_B(w)}{(z-w)^2}
+
\frac{\partial j_B(w)}{z-w}.
\label{eq:app-TjB-general}
\end{equation}
The mode commutator is obtained from
\begin{equation}
\bigl[
L_m^{\mathrm{tot}},
Q_B
\bigr]
=
\oint\frac{\mathrm{d}w}{2\pi i}
\oint_{z=w}\frac{\mathrm{d}z}{2\pi i}\,
z^{m+1}T^{\mathrm{tot}}(z)j_B(w).
\label{eq:app-Lm-QB-contour}
\end{equation}

For the fourth-order pole,
\begin{align}
\underset{z=w}{\operatorname{Res}}
\frac{z^{m+1}}{(z-w)^4}
&=
\frac{1}{3!}\,
\partial_w^3 w^{m+1}
\nonumber\\
&=
\frac{m(m^2-1)}{6}\,
w^{m-2},
\end{align}
so its contribution is
\begin{equation}
\frac{A}{6}m(m^2-1)c_m.
\end{equation}
For the third-order pole,
\begin{align}
\underset{z=w}{\operatorname{Res}}
\frac{z^{m+1}}{(z-w)^3}
&=
\frac{1}{2!}\,
\partial_w^2 w^{m+1}
\nonumber\\
&=
\frac{m(m+1)}{2}\,
w^{m-1},
\end{align}
and its contribution is
\begin{equation}
\frac{B}{2}m(m+1)s_m.
\end{equation}
Comparison with eq.~\eqref{eq:app-Lm-QB-anomaly} gives
\begin{equation}
A=\frac{c_{LL}^{\mathrm{tot}}}{2},
\qquad
B=c_{LS}^{\mathrm{tot}}.
\end{equation}
Therefore,
\begin{equation}
\begin{aligned}
T^{\mathrm{tot}}(z)j_B(w)
\sim{}&
\frac{c_{LL}^{\mathrm{tot}}}{2}
\frac{c(w)}{(z-w)^4}
+
c_{LS}^{\mathrm{tot}}
\frac{s(w)}{(z-w)^3}
\\
&+
\frac{j_B(w)}{(z-w)^2}
+
\frac{\partial j_B(w)}{z-w}.
\end{aligned}
\label{eq:app-TjB-CW}
\end{equation}

A primary field of conformal weight one obeys
\begin{equation}
T^{\mathrm{tot}}(z)\mathcal O(w)
\sim
\frac{\mathcal O(w)}{(z-w)^2}
+
\frac{\partial\mathcal O(w)}{z-w},
\label{eq:app-weight-one-primary}
\end{equation}
with no poles of order three or higher.
Equation~\eqref{eq:app-TjB-CW} therefore shows that \(j_B\) transforms as a
weight-one Witt tensor only if
\begin{equation}
c_{LL}^{\mathrm{tot}}=0,
\qquad
c_{LS}^{\mathrm{tot}}=0.
\label{eq:app-Witt-covariance-conditions}
\end{equation}
The fourth-order pole measures the Virasoro-type \((LL)\) obstruction, while
the third-order pole measures the mixed \((LS)\) obstruction.

\subsection*{Transformation under $S^{\mathrm{tot}}$.}

The \((SS)\) affine anomaly is not visible in the
\(T^{\mathrm{tot}}(z)j_B(w)\) OPE. It is detected by the transformation of
the BRST current under \(S^{\mathrm{tot}}\). The central part of the
corresponding mode commutator is
\begin{align}
\bigl[
S_m^{\mathrm{tot}},
Q_B
\bigr]_{\mathrm{anom}}
={}&
\sum_n c_{-n}
\bigl[
S_m^{\mathrm{tot}},
L_n^{\mathrm{tot}}
\bigr]_{\mathrm{cen}}
\nonumber\\
&+
\sum_n s_{-n}
\bigl[
S_m^{\mathrm{tot}},
S_n^{\mathrm{tot}}
\bigr]_{\mathrm{cen}}.
\label{eq:app-Sm-QB-start}
\end{align}
Antisymmetry of the mixed commutator gives
\begin{equation}
\bigl[
S_m^{\mathrm{tot}},
L_n^{\mathrm{tot}}
\bigr]_{\mathrm{cen}}
=
-\frac{c_{LS}^{\mathrm{tot}}}{2}
n(n+1)\delta_{m+n,0}.
\label{eq:app-SL-central}
\end{equation}
Setting \(n=-m\), one obtains
\begin{equation}
\bigl[
S_m^{\mathrm{tot}},
Q_B
\bigr]_{\mathrm{anom}}
=
-\frac{c_{LS}^{\mathrm{tot}}}{2}
m(m-1)c_m
+
c_{SS}^{\mathrm{tot}}\,m\,s_m.
\label{eq:app-Sm-QB-anomaly}
\end{equation}

Consider the anomalous part of the local OPE,
\begin{equation}
\left.
S^{\mathrm{tot}}(z)j_B(w)
\right|_{\mathrm{anom}}
\sim
\frac{C\,c(w)}{(z-w)^3}
+
\frac{E\,s(w)}{(z-w)^2}.
\label{eq:app-SjB-general}
\end{equation}
Since
\begin{equation}
S_m^{\mathrm{tot}}
=
\oint\frac{\mathrm{d}z}{2\pi i}\,
z^mS^{\mathrm{tot}}(z),
\end{equation}
the third-order pole contributes
\begin{equation}
\frac{C}{2}m(m-1)c_m,
\end{equation}
whereas the second-order pole contributes
\begin{equation}
E\,m\,s_m.
\end{equation}
Comparison with eq.~\eqref{eq:app-Sm-QB-anomaly} gives
\begin{equation}
C=-c_{LS}^{\mathrm{tot}},
\qquad
E=c_{SS}^{\mathrm{tot}}.
\end{equation}
Hence,
\begin{equation}
\left.
S^{\mathrm{tot}}(z)j_B(w)
\right|_{\mathrm{anom}}
\sim
-c_{LS}^{\mathrm{tot}}
\frac{c(w)}{(z-w)^3}
+
c_{SS}^{\mathrm{tot}}
\frac{s(w)}{(z-w)^2}.
\label{eq:app-SjB-CW}
\end{equation}
The absence of an anomalous transformation under \(S^{\mathrm{tot}}\)
therefore requires
\begin{equation}
c_{LS}^{\mathrm{tot}}=0,
\qquad
c_{SS}^{\mathrm{tot}}=0.
\label{eq:app-S-covariance-conditions}
\end{equation}

The \(M^{\mathrm{tot}}(z)j_B(w)\) OPE gives no additional anomaly condition
because the full quantum algebra contains no independent central extensions
in the \((LM)\), \((SM)\), or \((MM)\) channels.

\subsection*{Transformation under $M^{\mathrm{tot}}$.}

For completeness, consider the transformation under \(M^{\mathrm{tot}}\).
Its anomalous part is
\begin{align}
\bigl[M_m^{\mathrm{tot}},Q_B\bigr]_{\mathrm{anom}}
={}&
\sum_n c_{-n}
\bigl[M_m^{\mathrm{tot}},L_n^{\mathrm{tot}}\bigr]_{\mathrm{cen}}
+
\sum_n \widetilde c_{-n}
\bigl[M_m^{\mathrm{tot}},M_n^{\mathrm{tot}}\bigr]_{\mathrm{cen}}
\nonumber\\
&+
\sum_n s_{-n}
\bigl[M_m^{\mathrm{tot}},S_n^{\mathrm{tot}}\bigr]_{\mathrm{cen}}.
\end{align}
Since the quantum Carroll-Weyl algebra has no independent central
extensions in the \((LM)\), \((MM)\), or \((SM)\) channels,
\begin{equation}
\bigl[M_m^{\mathrm{tot}},Q_B\bigr]_{\mathrm{anom}}=0,
\qquad
\left.
M^{\mathrm{tot}}(z)j_B(w)
\right|_{\mathrm{anom}}=0.
\end{equation}
The noncentral terms cancel against the cubic ghost contributions by the
BFV construction. Thus the \(M^{\mathrm{tot}}j_B\) OPE provides a
consistency check but yields no additional anomaly condition.

\subsection*{Covariance conditions and the no-go result}

Combining eqs.~\eqref{eq:app-Witt-covariance-conditions} and
\eqref{eq:app-S-covariance-conditions}, covariance of the BRST current under
the full Carroll-Weyl quantum current algebra requires
\begin{equation}
c_{LL}^{\mathrm{tot}}
=
c_{LS}^{\mathrm{tot}}
=
c_{SS}^{\mathrm{tot}}
=
0.
\label{eq:app-BRST-current-covariance}
\end{equation}

For the matter system,
\begin{equation}
\bigl(
c_{LL},
c_{LS},
c_{SS}
\bigr)_{\mathrm{matter}}
=
(2D,-D,-D),
\label{eq:app-matter-vector}
\end{equation}
while the complete ghost system gives
\begin{equation}
\bigl(
c_{LL},
c_{LS},
c_{SS}
\bigr)_{\mathrm{ghost}}
=
(-54,6,4).
\label{eq:app-ghost-vector}
\end{equation}
Consequently,
\begin{equation}
c_{LL}^{\mathrm{tot}}=2D-54,
\qquad
c_{LS}^{\mathrm{tot}}=6-D,
\qquad
c_{SS}^{\mathrm{tot}}=4-D.
\label{eq:app-total-vector}
\end{equation}

The absence of the anomalous poles in eqs.~\eqref{eq:app-TjB-CW} and
\eqref{eq:app-SjB-CW} therefore requires
\begin{equation}
2D-54=0,
\qquad
6-D=0,
\qquad
4-D=0.
\label{eq:app-three-D-conditions}
\end{equation}
These equations respectively imply
\begin{equation}
D=27,
\qquad
D=6,
\qquad
D=4,
\end{equation}
and hence possess no common solution.

Up to the separate zero-mode and intercept conditions, the following
statements are therefore equivalent within the minimal highest-weight
complex,
\begin{equation}
\begin{aligned}
Q_B^2=0
&\quad\Longleftrightarrow\quad
c_{LL}^{\mathrm{tot}}
=
c_{LS}^{\mathrm{tot}}
=
c_{SS}^{\mathrm{tot}}
=
0
\\
&\quad\Longleftrightarrow\quad
\substack{
j_B\ \text{transforms as a weight-one tensor under }
T^{\mathrm{tot}},\\
\text{and transforms without anomalous terms under }
S^{\mathrm{tot}}.
}
\end{aligned}
\label{eq:app-equivalent-conditions}
\end{equation}

The BRST current consequently fails to transform covariantly under the full
quantum Carroll-Weyl current algebra in every target-space dimension. This
provides an alternative current-transformation derivation of the BRST
nilpotency obstruction obtained in the main text.

\section{Full two-coordinate Carrollian OPEs}
\label{app:two-coordinate}

\subsection*{Carrollian plane and evolution equations}

Introduce the BMS-plane coordinates
\begin{equation}
 x=\e^{\ii\sigma},\qquad y=\ii\tau\e^{\ii\sigma},
\qquad x_{12}=x_1-x_2,\qquad y_{12}=y_1-y_2.
\label{eq:BMS-plane}
\end{equation}
The mode algebra determines the second-coordinate dependence:
\begin{equation}
 \partial_yM=0,\qquad
 \partial_yT=\partial_xM,\qquad
 \partial_yS=-2M.
\label{eq:Carroll-evolution}
\end{equation}
Thus
\begin{equation}
 T(x,y)=T_0(x)+y\partial_xM(x),\qquad
 S(x,y)=S_0(x)-2yM(x).
\label{eq:current-lifts}
\end{equation}
The first equation is the standard BMS lift
\cite{Hao2022}; the second is fixed by \([S,M]=2M\).

\subsection*{Equal-time data and Carrollian evolution}

Eqs.~\eqref{eq:Carroll-evolution} show why the equal-time
calculation is sufficient to determine all central coefficients.
The \(y\)-dependence is polynomial and generated by \(M\), while the
central terms are \(c\)-numbers.  Starting with the reference-slice
currents \(T_0,M,S_0\),
\begin{equation}
 T=T_0+y\partial M,\qquad
 M=M,\qquad
 S=S_0-2yM,
\label{eq:lift-again}
\end{equation}
every unequal-\(y\) product is obtained by substitution and Taylor
expansion.  Since \(M M\) is regular and \(T M,S M\) have no central
poles, this evolution can generate descendants proportional to
\(y_{12}M\) but cannot create a new field-independent cocycle.

For example,
\begin{align}
 S(1)S(2)
 ={}&S_0(x_1)S_0(x_2)
 -2y_1M(x_1)S_0(x_2)
\nonumber\\
 &-2y_2S_0(x_1)M(x_2)
 +4y_1y_2M(x_1)M(x_2).
\label{eq:SS-lift-derivation}
\end{align}
The last term is regular.  Using the two orderings of the \(SM\) OPE
in the middle terms reorganizes their singular part into
\(4y_{12}M(2)/x_{12}\), while the equal-time affine pole remains
\(\cSS/x_{12}^2\).  This yields eq.~\eqref{eq:full-SS}.  The other lifted
products follow in the same way.

\subsection*{Mode vector fields and weights}

The BMS-plane coordinates arise from the exponential map of the
cylinder.  The centerless \(L,M\) generators act geometrically as
\begin{equation}
 \ell_n=-x^{n+1}\partial_x-(n+1)x^ny\partial_y,
 \qquad
 m_n=x^{n+1}\partial_y,
\label{eq:BMS-vector-fields}
\end{equation}
and obey
\begin{equation}
 [\ell_m,\ell_n]=(m-n)\ell_{m+n},\qquad
 [\ell_m,m_n]=(m-n)m_{m+n},\qquad
 [m_m,m_n]=0.
\label{eq:BMS-vector-algebra}
\end{equation}
The current \(S\) adds a weight-one tower whose adjoint action rescales
the \(M\) tower.  The fact that \(S\) has one lower conformal weight than
\(M\) explains both the third-order \(TS\) anomaly and the
second-order \(SS\) level.

The auxiliary spatial OPE is recovered by setting \(y_1=y_2=0\).
Conversely, lifting an equal-time OPE by
eq.~\eqref{eq:lift-again} is equivalent to evolving its modes with the
Carrollian Hamiltonian.  This establishes that the meromorphic
calculation in the main text is not discarding the \(y\)-dependence; it
uses it only after the independent singular data have been computed.

\subsection*{General centrally extended algebra}

For arbitrary \((\cL,\cS,\cSS)\), with the \(LM\) central charge set to
zero by the Jacobi identity, the full OPEs are
\begin{align}
 T(1)T(2)\sim{}&
 \frac{\cL}{2x_{12}^4}
 +\frac{2T(2)}{x_{12}^2}
 +\frac{\partial_xT(2)}{x_{12}}
 -\frac{4y_{12}M(2)}{x_{12}^3}
 -\frac{y_{12}\partial_yT(2)}{x_{12}^2},
\label{eq:full-TT}\\
 T(1)M(2)\sim{}&
 \frac{2M(2)}{x_{12}^2}
 +\frac{\partial_xM(2)}{x_{12}},
\label{eq:full-TM}\\
 T(1)S(2)\sim{}&
 \frac{\cS}{x_{12}^3}
 +\frac{S(2)}{x_{12}^2}
 +\frac{\partial_xS(2)}{x_{12}}
 -\frac{y_{12}\partial_yS(2)}{x_{12}^2},
\label{eq:full-TS}\\
 S(1)M(2)\sim{}&
 \frac{2M(2)}{x_{12}},
\label{eq:full-SM}\\
 S(1)S(2)\sim{}&
 \frac{\cSS}{x_{12}^2}
 +\frac{4y_{12}M(2)}{x_{12}},
\label{eq:full-SS}\\
 M(1)M(2)\sim{}&0.
\label{eq:full-MM}
\end{align}
All fields on the right-hand side are evaluated at point
\(2=(x_2,y_2)\).  At \(y_1=y_2=0\), these reduce to the equal-time OPEs
used in the main text.

\subsection*{Matter, ghosts, and total theory}

For matter,
\[
 (\cL,\cS,\cSS)_X=(2D,-D,-D);
\]
for ghosts,
\[
 (\cL,\cS,\cSS)_\gh=(-54,6,4);
\]
and for the total system,
\[
 (\cL,\cS,\cSS)_\tot=(2D-54,6-D,4-D).
\]
Substitution into
eqs.~\eqref{eq:full-TT}, \eqref{eq:full-TM}, \eqref{eq:full-TS}, \eqref{eq:full-SM}, \eqref{eq:full-SS}, and \eqref{eq:full-MM}
gives the complete two-coordinate Carrollian current algebra.  In
particular, the \(y_{12}\)-dependent terms are fixed descendants of the
equal-time algebra; they introduce no new independent central
coefficients and do not alter the nilpotency conditions.

\subsection*{Normalization table}

The abstract Weyl-BMS literature often uses a unit-charge dilatation
\(\mathcal D\) with \([\mathcal D,M]=M\), whereas the null-string
moment map is \(S=2\mathcal D\).  For clarity, all anomaly coefficients
in the two conventions are listed in \cref{tab:normalization-app}.

\begin{table}[H]
\centering
\renewcommand{\arraystretch}{1.14}
\begin{tabular}{|c|c|c|c|c|}
\hline
Sector & \(c_L\) &
\(c_{L\mathcal D}\) & \(c_{\mathcal D\mathcal D}\) &
\((\cS,\cSS)\) for \(S=2\mathcal D\)\\
\hline
Matter & \(2D\) & \(-D/2\) & \(-D/4\) & \((-D,-D)\)\\
Ghosts & \(-54\) & \(3\) & \(1\) & \((6,4)\)\\
Total & \(2D-54\) & \(3-D/2\) & \(1-D/4\)
& \((6-D,4-D)\)\\
\hline
\end{tabular}
\caption{Changing the normalization rescales mixed and affine
coefficients but leaves their zero sets unchanged.}
\label{tab:normalization-app}
\end{table}




\bibliographystyle{JHEP}
\bibliography{ref}

@article{Chen:2026cau,
    author = "Chen, Bin and Hu, Zezhou",
    title = "{Quantum Anomalies of Tensionless Bosonic Strings}",
    eprint = "2608.02987",
    archivePrefix = "arXiv",
    primaryClass = "hep-th",
    month = "8",
    year = "2026"
}

@article{Duary:2026bcs,
  author        = {Duary, Sarthak and Maji, Sourav},
  title         = {{Path integral quantization of null bosonic strings with Carroll-Weyl ghosts}},
  eprint        = {2606.04999},
  archivePrefix = {arXiv},
  primaryClass  = {hep-th},
  month         = {6},
  year          = {2026}
}

@article{Kato:1983,
  author  = {Kato, Mitsuhiro and Ogawa, Kaku},
  title   = {{Covariant quantization of string based on BRS invariance}},
  doi     = {10.1016/0550-3213(83)90680-6},
  journal = {Nucl. Phys. B},
  volume  = {212},
  pages   = {443--460},
  year    = {1983}
}

@book{Polchinski1,
    author = "Polchinski, J.",
    title = "{String theory. Vol. 1: An introduction to the bosonic string}",
    doi = "10.1017/CBO9780511816079",
    isbn = "978-0-511-25227-3, 978-0-521-67227-6, 978-0-521-63303-1",
    publisher = "Cambridge University Press",
    series = "Cambridge Monographs on Mathematical Physics",
    month = "12",
    year = "2007"
}

@article{Schild1977,
    author = "Schild, Alfred",
    title = "{Classical Null Strings}",
    reportNumber = "PRINT-76-0491 (TEXAS), ANL-HEP-PR-77-23",
    doi = "10.1103/PhysRevD.16.1722",
    journal = "Phys. Rev. D",
    volume = "16",
    pages = "1722",
    year = "1977"
}

@article{Isberg1994,
  author        = {Isberg, J. and Lindstrom, U. and Sundborg, B. and Theodoridis, G.},
  title         = {{Classical and quantized tensionless strings}},
  eprint        = {hep-th/9307108},
  archivePrefix = {arXiv},
  journal       = {Nucl. Phys. B},
  volume        = {411},
  pages         = {122--156},
  year          = {1994}
}

@article{Bagchi2013,
    author = "Bagchi, Arjun",
    title = "{Tensionless Strings and Galilean Conformal Algebra}",
    eprint = "1303.0291",
    archivePrefix = "arXiv",
    primaryClass = "hep-th",
    reportNumber = "MIT-CTP-4445, EMPG-13-02",
    doi = "10.1007/JHEP05(2013)141",
    journal = "JHEP",
    volume = "05",
    pages = "141",
    year = "2013"
}

@article{Lizzi1986,
    author = "Lizzi, F. and Rai, B. and Sparano, G. and Srivastava, A.",
    title = "{Quantization of the Null String and Absence of Critical Dimensions}",
    reportNumber = "RAL-86-086",
    doi = "10.1016/0370-2693(86)90101-2",
    journal = "Phys. Lett. B",
    volume = "182",
    pages = "326--330",
    year = "1986"
}

@article{Bozhilov1997,
  author        = {Bozhilov, P.},
  title         = {{Tensionless branes and the null string critical dimension}},
  eprint        = {hep-th/9711136},
  archivePrefix = {arXiv},
  journal       = {Mod. Phys. Lett. A},
  volume        = {13},
  pages         = {2571--2583},
  year          = {1998}
}

@article{BagchiVacua2021,
    author = "Bagchi, Arjun and Mandlik, Mangesh and Sharma, Punit",
    title = "{Tensionless tales: vacua and critical dimensions}",
    eprint = "2105.09682",
    archivePrefix = "arXiv",
    primaryClass = "hep-th",
    doi = "10.1007/JHEP08(2021)054",
    journal = "JHEP",
    volume = "08",
    pages = "054",
    year = "2021"
}

@article{BagchiReview2026,
    author = "Bagchi, Arjun and Banerjee, Aritra and Chatterjee, Ritankar and Pandit, Priyadarshini",
    title = "{The tensionless lives of null strings}",
    eprint = "2601.20959",
    archivePrefix = "arXiv",
    primaryClass = "hep-th",
    doi = "10.1016/j.physrep.2026.05.001",
    journal = "Phys. Rept.",
    volume = "1185",
    pages = "1--91",
    year = "2026"
}

@article{Chen2023,
    author = "Chen, Bin and Hu, Zezhou and Yu, Zhe-fei and Zheng, Yu-fan",
    title = "{Path-integral quantization of tensionless (super) string}",
    eprint = "2302.05975",
    archivePrefix = "arXiv",
    primaryClass = "hep-th",
    doi = "10.1007/JHEP08(2023)133",
    journal = "JHEP",
    volume = "08",
    pages = "133",
    year = "2023"
}

@article{Hao2022,
    author = "Hao, Peng-xiang and Song, Wei and Xie, Xianjin and Zhong, Yuan",
    title = "{BMS-invariant free scalar model}",
    eprint = "2111.04701",
    archivePrefix = "arXiv",
    primaryClass = "hep-th",
    doi = "10.1103/PhysRevD.105.125005",
    journal = "Phys. Rev. D",
    volume = "105",
    number = "12",
    pages = "125005",
    year = "2022"
}

@article{SheikhJabbari2026a,
  author        = {Sheikh-Jabbari, M. M. and Yavartanoo, H.},
  title         = {{On the consistency of null strings literature: The tale of an overlooked symmetry}},
  eprint        = {2605.12414},
  archivePrefix = {arXiv},
  primaryClass  = {hep-th},
  year          = {2026}
}

@article{SheikhJabbari2026b,
  author        = {Sheikh-Jabbari, M. M. and Yavartanoo, H.},
  title         = {{Null strings gauged and reloaded, I: Null strings have Carroll-Weyl gauge symmetry}},
  eprint        = {2605.25817},
  archivePrefix = {arXiv},
  primaryClass  = {hep-th},
  year          = {2026}
}

@article{SheikhJabbari2026c,
  author        = {Sheikh-Jabbari, M. M. and Yavartanoo, H.},
  title         = {{Null strings gauged and reloaded, II: Consistent classical treatment of the null strings}},
  eprint        = {2605.26822},
  archivePrefix = {arXiv},
  primaryClass  = {hep-th},
  year          = {2026}
}

@article{Batlle2020,
    author = "Batlle, Carles and Campello, V{\'\i}ctor and Gomis, Joaquim",
    title = "{A canonical realization of the Weyl BMS symmetry}",
    eprint = "2008.10290",
    archivePrefix = "arXiv",
    primaryClass = "hep-th",
    reportNumber = "ICCUB-20-017",
    doi = "10.1016/j.physletb.2020.135920",
    journal = "Phys. Lett. B",
    volume = "811",
    pages = "135920",
    year = "2020"
}

@article{FigueroaVishwa2025,
  author        = {Figueroa-O'Farrill, Jos{\'e} M. and Vishwa, Girish S.},
  title         = {{The BRST quantisation of chiral BMS-like field theories}},
  eprint        = {2407.12778},
  archivePrefix = {arXiv},
  primaryClass  = {hep-th},
  journal       = {J. Math. Phys.},
  volume        = {66},
  number        = {4},
  pages         = {042303},
  year          = {2025}
}

@article{Batlle2024,
  author        = {Batlle, Carles and Figueroa-O'Farrill, Jos{\'e} M. and Gomis, Joaquim and Vishwa, Girish S.},
  title         = {{BMS-like algebras: canonical realisations and BRST quantisation}},
  eprint        = {2411.14866},
  archivePrefix = {arXiv},
  primaryClass  = {hep-th},
  year          = {2024}
}

@article{Gustafsson1995,
  author        = {Gustafsson, H. and Lindstrom, U. and Saltsidis, P. and Sundborg, B. and van Unge, R.},
  title         = {{Hamiltonian BRST quantization of the conformal string}},
  eprint        = {hep-th/9410143},
  archivePrefix = {arXiv},
  journal       = {Nucl. Phys. B},
  volume        = {440},
  pages         = {495--520},
  year          = {1995}
}

@article{Rasulian2026,
  author        = {Rasulian, Ida M. and Sheikh-Jabbari, M. M. and Yavartanoo, H.},
  title         = {{Null-strings gauged, reloaded and quantized, I: Canonical quantization in the light-cone gauge}},
  eprint        = {2607.02970},
  archivePrefix = {arXiv},
  primaryClass  = {hep-th},
  year          = {2026}
}

@article{CasaliTourkine2016,
    author = "Casali, Eduardo and Tourkine, Piotr",
    title = "{On the null origin of the ambitwistor string}",
    eprint = "1606.05636",
    archivePrefix = "arXiv",
    primaryClass = "hep-th",
    doi = "10.1007/JHEP11(2016)036",
    journal = "JHEP",
    volume = "11",
    pages = "036",
    year = "2016"
}

@article{MasonSkinner2014,
    author = "Mason, Lionel and Skinner, David",
    title = "{Ambitwistor strings and the scattering equations}",
    eprint = "1311.2564",
    archivePrefix = "arXiv",
    primaryClass = "hep-th",
    doi = "10.1007/JHEP07(2014)048",
    journal = "JHEP",
    volume = "07",
    pages = "048",
    year = "2014"
}

@article{FigueroaHaveObers2025,
    author = "Figueroa-O'Farrill, Jos{\'e} and Have, Emil and Obers, Niels A.",
    title = "{Quantum carrollian bosonic strings}",
    eprint = "2509.04397",
    archivePrefix = "arXiv",
    primaryClass = "hep-th",
    reportNumber = "NORDITA 2025-048",
    doi = "10.1007/JHEP07(2026)098",
    journal = "JHEP",
    volume = "07",
    pages = "098",
    year = "2026"
}

@article{BagchiChakraborttyParekh2016,
  author        = {Bagchi, Arjun and Chakrabortty, Shankhadeep and Parekh, Pulastya},
  title         = {{Tensionless Strings from Worldsheet Symmetries}},
  eprint        = {1507.04361},
  archivePrefix = {arXiv},
  primaryClass  = {hep-th},
  journal       = {JHEP},
  volume        = {01},
  pages         = {158},
  doi           = {10.1007/JHEP01(2016)158},
  year          = {2016}
}

@article{CasaliHerfrayTourkine2017,
  author        = {Casali, Eduardo and Herfray, Yannick and Tourkine, Piotr},
  title         = {{The complex null string, Galilean conformal algebra and scattering equations}},
  eprint        = {1707.09900},
  archivePrefix = {arXiv},
  primaryClass  = {hep-th},
  journal       = {JHEP},
  volume        = {10},
  pages         = {164},
  doi           = {10.1007/JHEP10(2017)164},
  year          = {2017}
}

@article{Lindstrom:2026quz,
    author = {Lindstr{\"o}m, Ulf},
    title = "{Symmetries of tensionless strings}",
    eprint = "2605.26185",
    archivePrefix = "arXiv",
    primaryClass = "hep-th",
    reportNumber = "Uppsala Institute for Theoretical Physics preprint UUITP-08/26",
    month = "5",
    year = "2026"
}

@article{Lindstrom:2026zno,
    author = {Lindstr{\"o}m, Ulf},
    title = "{The conformal null string in $d+2$ and $d$ dimensions}",
    eprint = "2606.22498",
    archivePrefix = "arXiv",
    primaryClass = "hep-th",
    reportNumber = "Uppsala University, Theoretical Physics UUITP-15/26",
    month = "6",
    year = "2026"
}

@article{Bondi:1962px,
    author = "Bondi, H. and van der Burg, M. G. J. and Metzner, A. W. K.",
    title = "{Gravitational waves in general relativity. 7. Waves from axisymmetric isolated systems}",
    doi = "10.1098/rspa.1962.0161",
    journal = "Proc. Roy. Soc. Lond. A",
    volume = "269",
    pages = "21--52",
    year = "1962"
}

@article{Sachs1962AsymptoticSI,
    author = "Sachs, R.",
    title = "{Asymptotic symmetries in gravitational theory}",
    doi = "10.1103/PhysRev.128.2851",
    journal = "Phys. Rev.",
    volume = "128",
    pages = "2851--2864",
    year = "1962"
}

@article{Barnich:2006av,
    author = "Barnich, Glenn and Compere, Geoffrey",
    title = "{Classical central extension for asymptotic symmetries at null infinity in three spacetime dimensions}",
    eprint = "gr-qc/0610130",
    archivePrefix = "arXiv",
    reportNumber = "ULB-TH-06-08",
    doi = "10.1088/0264-9381/24/5/F01",
    journal = "Class. Quant. Grav.",
    volume = "24",
    pages = "F15--F23",
    year = "2007"
}

@article{Adami:2021nnf,
    author = "Adami, H. and Grumiller, D. and Sheikh-Jabbari, M. M. and Taghiloo, V. and Yavartanoo, H. and Zwikel, C.",
    title = "{Null boundary phase space: slicings, news {\&} memory}",
    eprint = "2110.04218",
    archivePrefix = "arXiv",
    primaryClass = "hep-th",
    doi = "10.1007/JHEP11(2021)155",
    journal = "JHEP",
    volume = "11",
    pages = "155",
    year = "2021"
}

@article{Adami:2020ugu,
    author = "Adami, H. and Sheikh-Jabbari, M. M. and Taghiloo, V. and Yavartanoo, H. and Zwikel, C.",
    title = "{Symmetries at null boundaries: two and three dimensional gravity cases}",
    eprint = "2007.12759",
    archivePrefix = "arXiv",
    primaryClass = "hep-th",
    doi = "10.1007/JHEP10(2020)107",
    journal = "JHEP",
    volume = "10",
    pages = "107",
    year = "2020"
}

@article{Ballesteros:2026bqe,
    author = "Ballesteros, Romina and Lescano, Eric and Pati{\~n}o-L{\'o}pez, Sergio",
    title = "{Carrollian limit of NS-NS and Heterotic Supergravity}",
    eprint = "2607.09847",
    archivePrefix = "arXiv",
    primaryClass = "hep-th",
    month = "7",
    year = "2026"
}

@article{Lescano:2026hxd,
    author = "Lescano, Eric",
    title = "{Carrollian bosonic supergravity at order $α'$ and the universal cancellation of higher-curvature divergences}",
    eprint = "2607.24916",
    archivePrefix = "arXiv",
    primaryClass = "hep-th",
    month = "7",
    year = "2026"
}

@article{DuaryMajiADS,
    author = "Duary, Sarthak and Maji, Sourav",
    title = "{Carroll-Weyl gauged null strings in AdS}",
    eprint = "2610.10717",
    archivePrefix = "arXiv",
    primaryClass = "hep-th",
    month = "10",
    year = "2026"
}

@article{DuaryMajiNullStringHolography,
    author = "Duary, Sarthak and Maji, Sourav",
    title = "{Null string holography in AdS from local BRST descent}",
    eprint = "2610.10540",
    archivePrefix = "arXiv",
    primaryClass = "hep-th",
    month = "10",
    year = "2026"
}

\end{document}